\documentclass[11pt]{article}

\usepackage[utf8]{inputenc}
\usepackage{hyperref}
\usepackage[square, numbers]{natbib}

\usepackage[english]{babel}
\usepackage{mathtools}
\usepackage{amsmath}
\usepackage{amsfonts}
\usepackage{verbatim}
\usepackage[title,titletoc,toc]{appendix}

\usepackage[section]{placeins}

\usepackage{geometry}
\usepackage{booktabs,array}
\makeatletter
\newcolumntype{V}[1]{>{\topsep=0pt\@minipagetrue}p{#1}<{\vspace{-\baselineskip}}}
\makeatother

\newcommand{\subpanelref}[2]{\hyperref[#1]{\ref*{#1}#2}}
\newcommand {\real} {\mathbb{R}}
\newcommand{\rd}{\text{\upshape d}}
\newcommand{\ba}{\ensuremath{\mathbf{a}}}

\newcommand{\bx}{\ensuremath{\mathbf{x}}}
\newcommand{\by}{\ensuremath{\mathbf{y}}}
\newcommand {\bGamma} {\boldsymbol{\Gamma}}

\newcommand {\btheta} {\mbox{\boldmath $\theta$}}

\newcommand{\cD}{\ensuremath{\mathcal{D}}}
\newcommand{\cM}{\ensuremath{\mathcal{M}}}
\newcommand{\cN}{\ensuremath{\mathcal{N}}}

\newcommand{\fM}{\ensuremath{\mathfrak{M}}}

\title{Increasing certainty in systems biology models using Bayesian multimodel inference}

\author
{Nathaniel Linden-Santangeli,$^{1}$ Jin Zhang,$^{2}$ Boris Kramer,$^{1\ast}$ Padmini Rangamani$^{1\ast}$\\
\\
\normalsize{$^{1}$Department of Mechanical and Aerospace Engineering, University of California San Diego,}\\
\normalsize{9500 Gilman Dr, La Jolla, CA 92093, USA}\\
\normalsize{$^{2}$Department of Pharmacology, University of California San Diego,}\\
\normalsize{9500 Gilman Dr, La Jolla, CA 92093, USA}\\
\\
\normalsize{$^\ast$To whom correspondence should be addressed; E-mail:  bmkramer@ucsd.edu}; \\ \normalsize{prangamani@ucsd.edu.}
}

\date{}

\begin{document} 
\maketitle 

\section*{Abstract}
Mathematical models are indispensable to the system biology toolkit for studying the structure and behavior of intracellular signaling networks.
A common approach to modeling is to develop a system of equations that encode the known biology using approximations and simplifying assumptions.
As a result, the same signaling pathway can be represented by multiple models, each with its set of underlying assumptions, which opens up challenges for model selection and decreases certainty in model predictions. 
Here, we use Bayesian multimodel inference to develop a framework to increase certainty in systems biology models.
Using models of the extracellular regulated kinase (ERK) pathway, we first show that multimodel inference increases predictive certainty and yields predictors that are robust to changes in the set of available models.
We then show that predictions made with multimodel inference are robust to data uncertainties introduced by decreasing the measurement duration and reducing the sample size.
Finally, we use multimodel inference to identify a new model to explain experimentally measured sub-cellular location-specific ERK activity dynamics.
In summary, our framework highlights multimodel inference as a disciplined approach to increasing the certainty of intracellular signaling activity predictions.

%%%%%%%%%%%%%%%%%%%%%%%%%%%%%%%%%%%%%%%%%%%%%%%%%%%%%%%%%%%%%%%
% INTRODUCTION
%%%%%%%%%%%%%%%%%%%%%%%%%%%%%%%%%%%%%%%%%%%%%%%%%%%%%%%%%%%%%%%
\section{Introduction} \label{sec:intro}

Current innovations in molecular tools~\cite{greenwaldGeneticallyEncodedFluorescent2018, freiNextGenerationGeneticallyEncoded2024} and high-resolution microscopy~\cite{schermellehSuperresolutionMicroscopyDemystified2019,stoneSuperResolutionMicroscopyShedding2017} have led to new discoveries in cellular signal transduction including spatial regulation of signaling pathways.
These discoveries require new mathematical models to give rise to mechanistic insights.
Furthermore, mathematical models also enable the generation of experimentally testable predictions of intracellular signaling~\cite{bhallaEmergentPropertiesNetworks1999,Kitano2002-vh, Kholodenko2010-ta}.
However, one key challenge in systems biology is formulating a model when there are many unknowns.
As a result, the same signaling pathway can be described by different mathematical models that vary in their simplifying assumptions and model formulations.
For example, searching the popular BioModels database for models of the extracellular-regulated kinase signaling cascade (ERK) yields over 125 results for models that use ordinary differential equations~\cite{BioModels2018a,BioModels2020}.
While all of these models likely apply to certain scenarios, it is unclear how one might select a model given certain experimental observations.
In this work, we explore two important questions:
(1) \textit{How can we quantify the effects of uncertainty in the model formulation, called model uncertainty, on model predictions?} and 
(2) \textit{How can we reduce model choice uncertainty to increase the certainty of intracellular signaling predictions?}
Here, using ERK signaling as our model system, we show that Bayesian multimodel inference (MMI) increases predictive certainty by leveraging the available data and accounting for all of the user-specified models.

The goal of uncertainty quantification in systems biology is to understand how model assumptions, inferred quantities, and data uncertainties impact model predictions~\cite{Smith2013-lw, Geris2016-xv}.
Bayesian parameter estimation quantitatively assesses parametric uncertainty by estimating a probability distribution for unknown parameters, such as reaction rate constants and equilibrium coefficients, from training data~\cite{Linden2022-xi, Smith2013-lw, Gelman2014-jj}.
However, at the model level, explicit approaches to handle model uncertainty are often not employed in systems biology.
Model selection using information criteria~\cite{Burnham2002-rl, Akaike1974-cp} or Bayes Factors~\cite{Raftery2010-nx} has been the preferred approach in systems biology to select a single ``best'' model when multiple models are available~\cite{Kirk2013-pn, Burnham2002-rl}.
However, given the limited and noisy data often available in systems biology, these approaches may limit predictive performance by introducing selection biases and misrepresentations of uncertainty ~\cite{Burnham2002-rl,Vehtari2016-sb}.

Multimodel inference avoids selection biases and accounts for model uncertainty by including contributions from every specified model~\cite{Burnham2002-rl,Vehtari2016-sb,Gelman2014-jj,Hoeting1999-hs,Stumpf2006-ia,Stumpf2020-em}.
Theoretical results have shown that MMI can improve predictive performance by reducing uncertainty and increasing robustness to modeling assumptions~\cite{Yao2018-gn,Vehtari2016-sb,Burnham2002-rl,Clemen1999-my,Winkler1981-jw}.
MMI  methods combine model predictions by taking a weighted average over all supplied models with weights chosen according to a specified criterion~\cite{Stone1961-vw, Winkler1981-jw,Clemen1999-my}.
Methods for choosing the weights range from parametric consensus estimation~\cite{Winkler1981-jw, Clemen1999-my} and frequentist Akaike-weighing~\cite{Burnham2002-rl} to Bayesian model averaging (BMA)~\cite{Hoeting1999-hs}, pseudo-Bayesian model averaging (pseudo-BMA)~\cite{Vehtari2016-sb,Yao2018-gn}, and stacking of predictive densities (stacking)~\cite{Yao2018-gn}.

Previous applications of MMI to systems biology have focused on a limited subset of the available methods, primarily the information-criterion-based approach and Bayesian model averaging.
Stumpf et al.\ conducted a theoretical analysis of MMI for biological network inference~\cite{Stumpf2006-ia, Stumpf2020-em}.
Based on these analyses and limited applications to protein-protein interaction data, the authors concluded that it is generally important to choose a good set of models for building accurate MMI estimates.
More recently, Beik et al.\ utilized MMI with BMA to select candidate tumor growth mechanisms that are consistent with several experimental datasets~\cite{Beik2023-fs}.
However, MMI has not yet been investigated as a method to increase the certainty of intracellular signaling predictions in systems biology.

In this work, we develop a framework for increasing certainty in systems biology models by employing Bayesian multimodel inference for ERK signaling.
Specifically, we investigate how MMI can account for model uncertainty, increase predictive certainty, and aid in model selection using established models of ERK signaling from the literature.
We select ten popular  ERK signaling models and estimate the kinetic parameters from synthetic and experimental data with Bayesian inference.
Using synthetic data, we first show that MMI increases the certainty of both EGF-ERK dose-response curve predictions and time-dependent ERK activity predictions (Figure~\ref{fig:synth-DR}).
Next, we show that MMI generates predictions that are robust to changes in the composition of the set of supplied ERK models (Figure~\ref{fig:synth-mmi-DR-perturb}) and to increasing uncertainty in the data (Figure~\ref{fig:keyes-qualityCYTO}). 
We then apply MMI to study the mechanisms driving sub-cellular location-specific ERK activity observed by Keyes et al.~\cite{Keyes2020-ub}.
We find that Bayesian parameter estimation alone is not sufficient to capture the mechanism of sub-cellular variability in ERK signaling; importantly, Bayesian MMI predicts that location-specific differences in both Rap1 activation and negative feedback strength are necessary to capture the observed dynamics (Figure~\ref{fig:keyes-Rap1-NegFeed}).
We conclude that MMI increases predictive certainty when multiple models of the same signaling pathway are available via a structured approach to simultaneously handle model uncertainty and model selection.

%%%%%%%%%%%%%%%%%%%%%%%%%%%%%%%%%%%%%%%%%%%%%%%%%%%%%%%%%%%%%%%
% RESULTS
%%%%%%%%%%%%%%%%%%%%%%%%%%%%%%%%%%%%%%%%%%%%%%%%%%%%%%%%%%%%%%%
\section{Results} \label{sec:results}

\subsection{Bayesian multimodel inference combines user-specified models to account for model uncertainty} \label{sec:mmi}

\begin{figure}[h!]
    \centering
    \includegraphics[width=\textwidth]{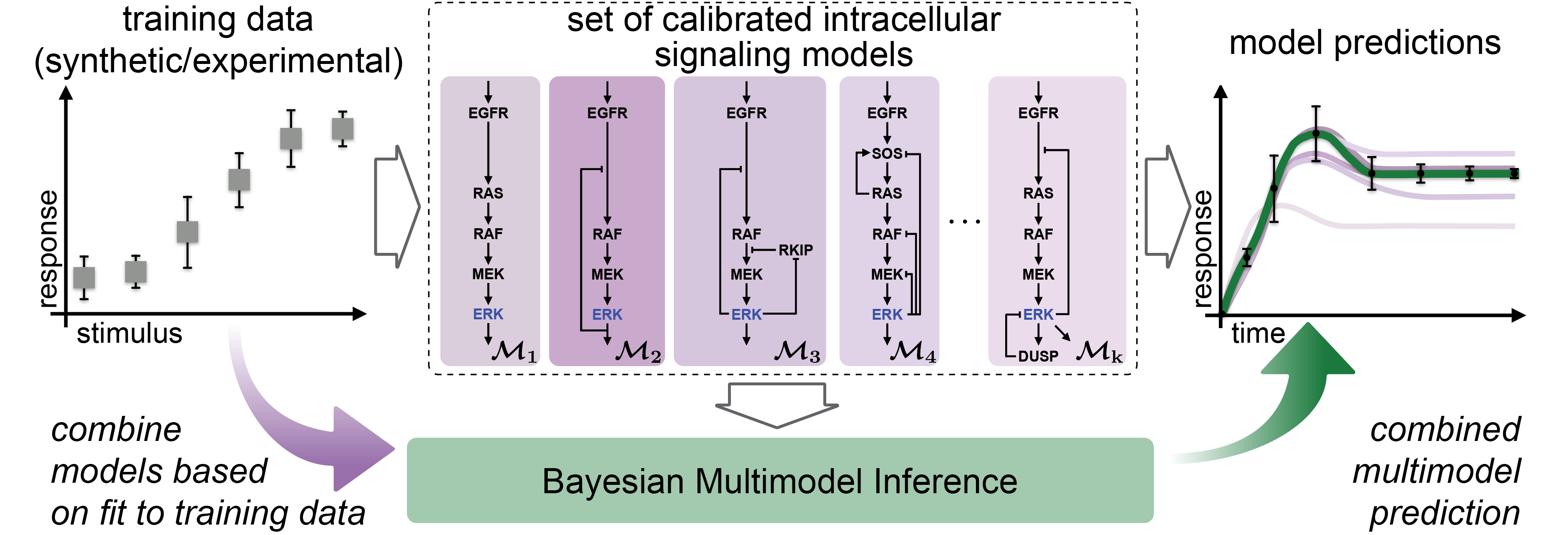}
    \caption{\textbf{Bayesian multimodel inference accounts for model uncertainty by incorporating contributions from intracellular signaling models specified by the user.}
    Using ten models of extracellular-regulated kinase signaling (ERK) as our focus, we outline the steps in the Bayesian multimodel inference (MMI) framework. 
    First, we use Bayesian parameter estimation to calibrate the parameters in each model in the set to the training data.
    Each calibrated model can then be used to predict biological quantities; however, the accuracy and uncertainty of each model's predictions may vary.
    Next, Bayesian MMI constructs a new estimator by combining specified models based on how well they predict the training data.
    Finally, the multimodel inference-based estimator improves predictive accuracy and robustness compared to individual models.
    }
    \label{fig:MMI-overview}
\end{figure}

Bayesian multimodel inference systematically constructs a new consensus estimator of important systems biology \textit{quantities of interest} (QoIs) that accounts for model uncertainty.
For ERK signaling, the QoI, $q$, is either the EGF-ERK dose-response curve ($q(u_i)$ where $u_i = [\text{EGF}]$) or the time-dependent trajectory of EGF-induced ERK activity ($q(t)$ at time $t$).
The framework for Bayesian multimodel inference is shown in Figure~\ref{fig:MMI-overview}, where we (1) calibrate available models to training data with Bayesian inference, (2) combine the resulting predictive probability densities using MMI, and (3) provide improved multimodel predictions of important quantities in systems biology studies.

The goal of Bayesian MMI is to build a multimodel estimate of the QoI, defined as ${\rm p}(q|\cD_{\rm train}, \fM_K)$, that leverages the entire set of specified models, $\fM_K = \{\cM_1, \dots, \cM_K\}$~\cite{Geweke2011-mf, Yao2018-gn, Hoeting1999-hs, Burnham2002-rl}.
Notably, we leverage the Bayesian framework to fully characterize predictive uncertainty through predictive probability densities~\cite{Gelman2014-jj,Smith2013-lw}.
The training data, $\cD_{\rm train} = \{\by^1, \ldots, \by^{D_{\rm train}}\}$, consists of $D_{\rm train}$ noisy experimental observations (or synthetic data), and can correspond to time points $t^i$ in dynamic responses $\by^i = \by(t^i)$ or to input stimuli $u^i$ in dose-response curves $\by^i = \by(u^i)$.
In this work, we use Bayesian methods to estimate unknown model parameters from training data.
Thus, after parameter estimation, each model predicts a probability density for the QoI, ${\rm p}(\hat{q}_k)$~\cite{Linden2022-xi, Gelman1998-vb}.
Bayesian MMI then revolves around combining the predictive densities from each model into a single multimodel predictive density.

A standard approach for MMI is to take a linear combination of predictive densities from each model,
\begin{equation*}
    {\rm p}(q|\cD_{\rm train}, \fM_K) \coloneqq \sum_{k=1}^K w_k {\rm p}(q_k | \cM_k, \cD_{\rm train}), \label{eq:mod-avg}
\end{equation*}
with weights $w_k \geq 0$ and $\sum_k^K w_k = 1$~\cite{Hoeting1999-hs, Burnham2002-rl, Martin2021-ly, Stone1961-vw, Geweke2011-mf}.
We note that the weights can either be scalars or realizations of a probability mass function defined over the set of models.
The key challenge is estimating the weight to assign each predictive density.
In this work, we compare three methods for choosing the weights, $w_k$: \textit{Bayesian model averaging} (BMA)~\cite{Hoeting1999-hs}, \textit{pseudo-Bayesian model averaging} (pseudo-BMA)~\cite{Geweke2011-mf, Yao2018-gn}, and \textit{stacking of predictive densities} (stacking)~\cite{Yao2018-gn}.
Importantly, the potential for MMI to bring new insights into intracellular signaling has not previously been explored, and because each method for MMI has distinct advantages and disadvantages, we chose to compare all three.
We briefly summarize the methods below. 

Bayesian model averaging weighs each model by the model probability conditioned on the training data, $w_k^{\rm BMA}={\rm p}(\cM_k|\cD_{\rm train})$~\cite{Hoeting1999-hs}.
The model probability quantifies the probability of model $\cM_k$ correctly predicting the training data relative to the other models in the set.
While BMA is the natural Bayesian approach to MMI, the method suffers from several key challenges, including the necessary computation of the marginal likelihood, strong dependence on prior information, and reliance on data-fit alone instead of on predictive performance~\cite{Clyde1999-kq, Hoeting1999-hs, Gelman2014-jj, Yao2018-gn}.
Due to these potential challenges of BMA, we also investigate pseudo-BMA and stacking for MMI of ERK signaling.

Pseudo-Bayesian model averaging assigns model weights based on the expected future predictive performance measured with the \textit{expected log pointwise predictive density} (ELPD)~\cite{Vehtari2016-sb, Yao2018-gn}.
The ELPD quantifies the expected predictive performance of a model on new data by computing the distance between the predictive density and the true data-generating density.
However, the ELPD is intractable directly because we do not know the true data-generating density, so we instead estimate the ELPD with Pareto smoothed importance sampling leave-one-out cross-validation (PSIS-LOO-CV)~\cite{Vehtari2016-sb}.
Pseudo-BMA normalizes the estimated ELPD of each model, $\widehat{{\rm ELPD}}_k^{\rm LOO}$, to the sum of that quantity across all models to construct model weights, $w_k^{\rm pBMA}$.
The two key challenges of pseudo-BMA are that it relies on potentially erroneous estimates of the ELPD, such as PSIS-LOO-CV, and that pseudo-BMA weights do not account for correlations between individual model predictions~\cite{Yao2018-gn}.

Stacking of predictive densities selects optimal model weights, $w_k^{\rm stack}$, to maximize the ELPD of the consensus density.
We follow the approach introduced in~\cite{Yao2018-gn}, which maximizes the log score between the consensus predictive density and the true data-generating distribution estimated by the ELPD.
In contrast to BMA and pseudo-BMA, which weigh models independently, stacking weighs all models simultaneously~\cite{Martin2021-ly}.
Thus, stacking can be shown to find the best-estimating density that is closest (in terms of the log scoring rule used to define the optimality criterion) to the data-generating process~\cite{Yao2018-gn}.
However, similar to pseudo-BMA, stacking relies on ELPD estimates for MMI.
Based on each method's potential advantages and disadvantages, the proposed MMI framework allows any MMI method to be used, enabling us to compare the methods in the context of intracellular signaling.

\subsection{Set of ERK signaling models highlights model uncertainty due to the variations in model formulation}\label{sec:ERK-family}

Extracellular-regulated kinase signaling plays a key role in controlling a number of cellular processes including proliferation, growth, metabolism, and differentiation~\cite{roskoskiERK1MAPKinases2012, lavoieERKSignallingMaster2020}.
Due to this widespread importance, ERK is one of the most extensively modeled intracellular signaling pathways~\cite{Orton2005-uo}.
As a test problem for MMI, we specified a set of ten models of the core ERK signaling network~\cite{Huang1996-ki,Levchenko2000-gg,Kholodenko2000-ot,Hornberg2005-fs,Birtwistle2007-dw,Orton2008-lm,Von_Kriegsheim2009-nq,Shin2014-nd,Ryu2015-ix,Kochanczyk2017-jc}.
We selected these ten models because they all span from EGF-receptor binding and activation to ERK activation and because the model equations were readily available from original or secondary sources (see Supplemental Table~\ref{tab:ERK-model-info} and Supplemental Materials for more details).
The models also only focus on the core ERK kinase cascade without including additional crosstalk with other signaling pathways.
While the models are similar in scope, they vary in the assumed biological complexity, the feedback mechanisms included, and the mathematical formulation used to represent the reaction kinetics. 
For example, the model from~\cite{Huang1996-ki}, H'~1996, includes no feedback loops and uses mass action kinetics, while that from~\cite{Ryu2015-ix}, R'~2015, includes two negative feedback loops and uses normalized Hill equations.
Throughout this work, we refer to each model by the first letter of the first author's last name and the publication year, e.g., Huang and Ferrell 1996~\cite{Huang1996-ki} is H'~1996.
The large variations in model formulation are reflected by the range in the number of state variables and parameters included in each model.
Thus, we conclude that the curated set of ERK signaling models is a good example of model uncertainty because the ten models are similar in scope but vary in mathematical formulation.

We previously showed that \textit{a priori} structural identifiability and global sensitivity analyses are critical to successful Bayesian parameter estimation for intracellular signaling models~\cite{Linden2022-xi}.
Therefore, we performed local structural identifiability analysis on each model to determine which parameters can be uniquely identified locally in parameter space.
Next, we performed global sensitivity analysis on each model using Morris screening~\cite{saltelliGlobalSensitivityAnalysis2007} to determine which identifiable parameters significantly influence predicted ERK activity (Supplemental Figure~\ref{sup-fig:gsa}).
Based on the results of both analyses, we reduced the number of free parameters by fixing all nonidentifiable and noninfluential parameters to nominal values from the literature and only estimated the remaining parameters (see Supplemental Materials).

\subsection{Bayesian Multimodel inference provides a structured approach to increase model certainty in predictions of EGF-induced ERK activity} \label{sec:mmi-synth}

\begin{figure}[hb!]
    \centering
    \includegraphics[width=\textwidth]{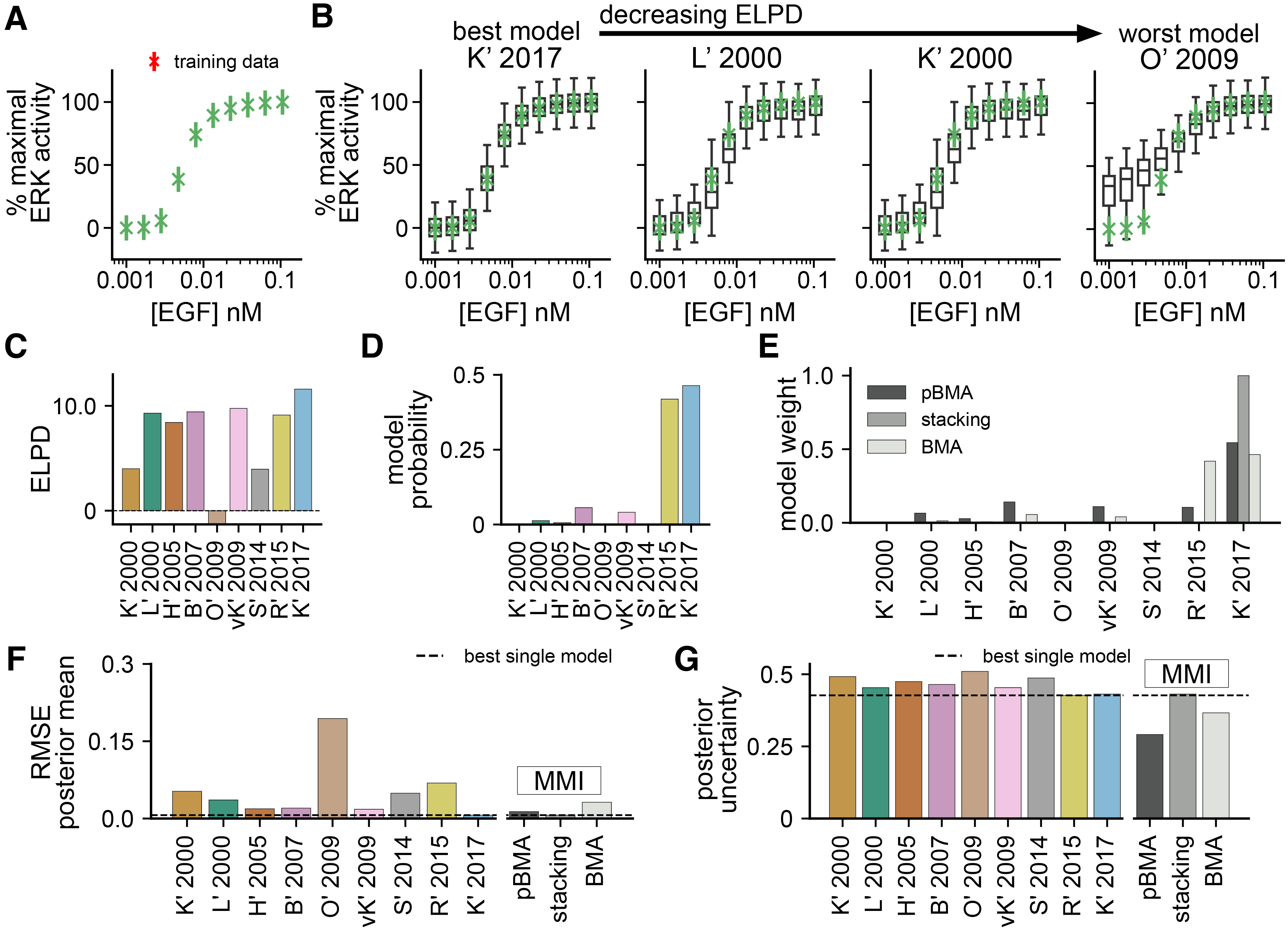}
    \caption{\textbf{Multimodel inference reduces uncertainty in EGF-ERK dose-response curve predictions.}
    (\textbf{A}) Synthetic EGF-ERK dose-response curve, generated by stimulating the H'~1996 model~\cite{Huang1996-ki} to steady-state at ten EGF input levels ranging from $0.001$--$0.106$ nM and computing the percentage of active ERK at each EGF stimulus level relative to the maximum across all stimuli.
    Data are assumed to have Gaussian measurement noise with a standard deviation of $\sigma = 0.1$.
    (\textbf{B}) 
    Posterior predictive densities of the EGF-ERK dose-response for four out of nine models ordered by decreasing expected log pointwise predictive density (ELPD).
    Additional densities are shown in Supplemental Figure~\ref{sup-fig:synth-DR-post-pred}.
    (\textbf{C}) 
    ELPD values for all models computed using Pareto smoothed importance sampling leave-one-out cross-validation (PSIS-LOO-CV).
    Models with higher relative ELPD are more likely to predict future data correctly.
    Estimated ELPD values are used to construct MMI estimates with pseudo-Bayesian model averaging (pseudo-BMA) and stacking of predictive densities (stacking).
    (\textbf{D}) 
    Model probabilities are computed using sequential Monte Carlo estimates of the marginal likelihood.
    The model probability is the probability of each model conditioned on the training data and is used for MMI with Bayesian model averaging (BMA).
    (\textbf{E})
    MMI model weights for all models using pseudo-BMA, BMA, and stacking.
    (\textbf{F}) Root mean square error (RMSE) of the posterior mean dose-response prediction for each model and the multimodel predictions.
    The dashed black line shows the lowest RMSE of any single model.
    (\textbf{G}) Posterior uncertainty measured by the mean $95\%$ credible interval width (the interval between $2.5^{\rm th}$ and $97.5^{\rm th}$ percentiles) over all EGF levels.
    The dashed black line shows the lowest uncertainty of any single model.}
    \label{fig:synth-DR}
\end{figure}

As a first test, we applied MMI to predict EGF-induced ERK activity from synthetic training data.
To do so, we separately estimated the parameters of nine out the ten models, excluding H'~1996, to sets of synthetic EGF-ERK dose-response training data that we generated with the H'~1996 model~\cite{Huang1996-ki} (Figure~\subpanelref{fig:synth-DR}{A}).
Using the resulting posterior samples, we constructed the posterior predictive densities for the EGF-ERK dose-response curve using each model separately (Figure~\subpanelref{fig:synth-DR}{B} and Supplemental Figure~\subpanelref{sup-fig:synth-DR-post-pred}{A}).
In the dose-response predictions, we observe that several models, L'~2000, H'~2005, B'~2007, S'~2014, R'~2015, and K'~2017, qualitatively predict the dose-response curve equally well.
The remaining models, K'~2000 and O'~2009, introduce errors, such as a decreased dynamic range or greater predictive uncertainty.
We attribute the observed variations in the dose-response predictions to variations in the formulation of each model.

Each of the three MMI methods distributes MMI weights differently between the models, but the MMI predictions appear similar.
The exact ranking of the models using the ELDP values (Figure~\subpanelref{fig:synth-DR}{C}) and the model probabilities (Figure~\subpanelref{fig:synth-DR}{D}) varies between the two methods, despite the overall trends remaining similar.
In general, the models that predict the dose-response curve qualitatively better receive larger ELPD values and higher model probability than those that do so poorly.
Both pseudo-BMA and BMA assign non-zero weights to more than one model, but they weigh different subsets of the model family (Figure~\subpanelref{fig:synth-DR}{E}).
Meanwhile, stacking places all weight on the K'~2017 model.
Despite the different weight distributions, each of the three MMI methods predicts dose-response curves that all appear qualitatively similar to the data (Supplemental Figure~\subpanelref{sup-fig:synth-DR-post-pred}{B}--\subpanelref{sup-fig:synth-DR-post-pred}{D}).

Quantitative differences in the predictions for each of the methods show that only pseudo-BMA and BMA led to increased predictive certainty.
The \textit{root mean square error} (RMSE) of the posterior mean of the dose-response curve is lower than that of all but one model for pseudo-BMA and stacking, but not for BMA (Figure~\subpanelref{fig:synth-DR}{F}).
Furthermore, the predictive uncertainty measured by the averaged width of the posterior predictive $95\%$ credible interval is substantially lower than any individual model for both pseudo-BMA and BMA, but not for stacking (Figure~\subpanelref{fig:synth-DR}{G}).
Notably, we observe this uncertainty reduction for the two methods that weighted average over multiple models---pseudo-BMA and BMA---but not for stacking, which only selects a single model.
These results show that MMI predictions retain the predictive accuracy of the best models and can also yield reductions in uncertainty when a weighted average is used.
From this, we conclude that MMI is able to increase predictive certainty by combining multiple models while retaining the accuracy of the best models.

Furthermore, the ability of MMI to reduce predictive uncertainty is not limited to dose-response data.
To test this, we repeated MMI for the same set of models using synthetic EGF-induced dynamic ERK activity data (Supplemental Figure~\subpanelref{sup-fig:synth-traj}{A}).
Similar to the previous dose-response case, we observe that predictions vary between individual models (Supplemental Figure~\subpanelref{sup-fig:synth-traj}{B}~and~\subpanelref{sup-fig:synth-traj-post-pred}{A}).
Further, we again find that the different MMI methods yield different weight assignments (Supplemental Figure~\subpanelref{sup-fig:synth-traj}{E}), and, thus, different MMI predictions (Supplemental Figure~\subpanelref{sup-fig:synth-traj-post-pred}{B}).
Interestingly, the best model in this case was vK'~2009, which was different from the one for the dose-response predictions.
The predictive error of the MMI estimates was again on par with the models with the lowest error (Supplemental Figure~\subpanelref{sup-fig:synth-traj}{F}), and MMI again reduced posterior uncertainty when multiple models are weighted averaged (Supplemental Figure~\subpanelref{sup-fig:synth-traj}{G}).
Therefore, we conclude that MMI handles model uncertainty in predictions of both the EGF-ERK dose-response curve and EGF-induced dynamic ERK activity.
Importantly, in both contexts, MMI predictions had lower errors than choosing a model at random and showed reduced uncertainty when using pseudo-BMA and BMA. 
This suggests that MMI can increase predictive certainty in multiple contexts of intracellular signaling.

\subsection{Multimodel inference improves predictions of ERK activity by increasing robustness to model uncertainty} \label{sec:mmi-model-robust}

To understand how increasing model uncertainty affects MMI predictions, we systematically varied the composition of the model set.
Specifically, we asked how (1) adding a bad model, (2) removing the best model, and (3) changing the size of the model set affects MMI predictive performance.

\begin{figure}[hb!]
    \centering
    \includegraphics[width=\textwidth]{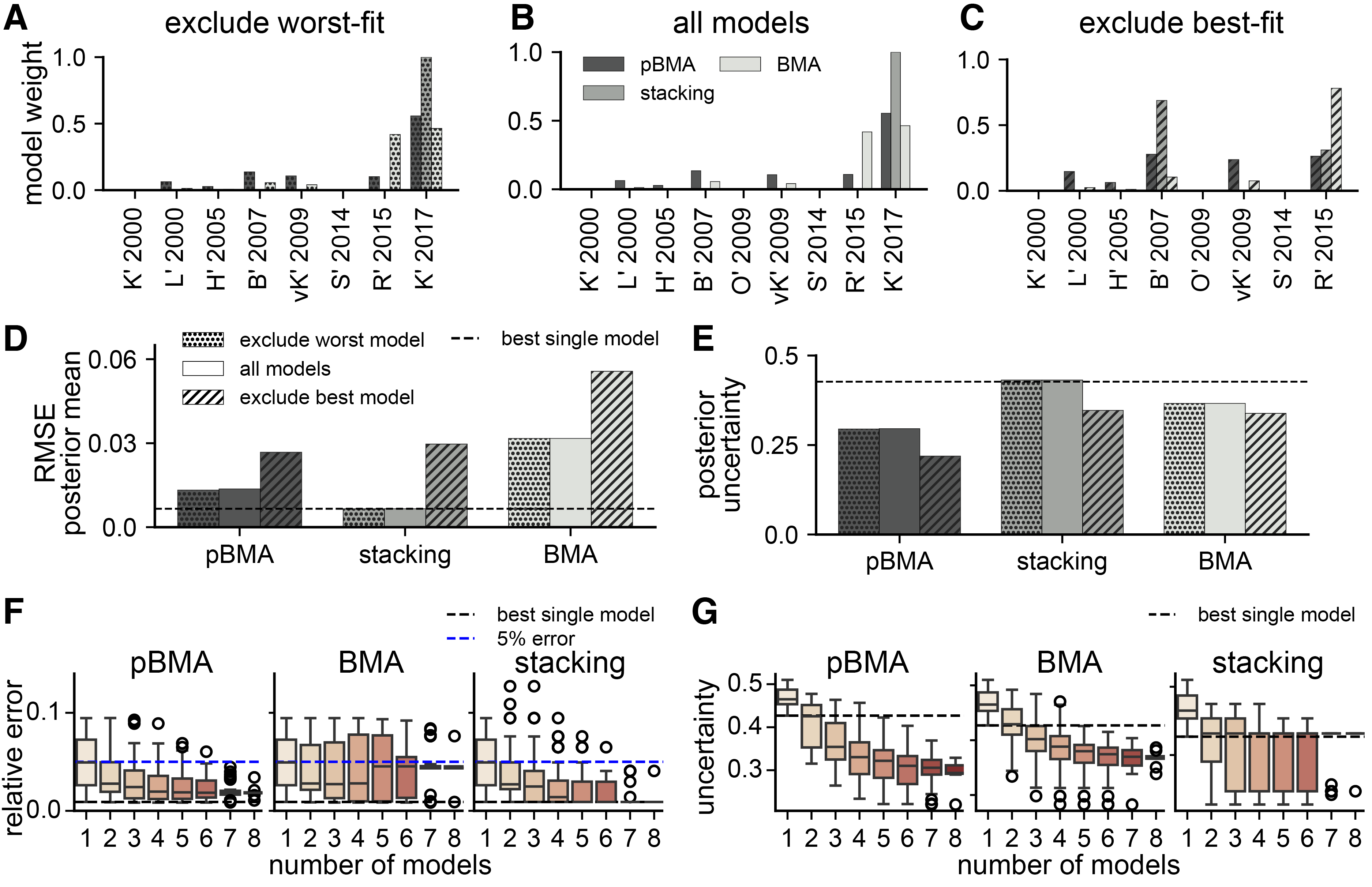}
    \caption{
        \textbf{Bayesian multimodel inference is robust to perturbations in the set of plausible models for EGF-ERK dose-response predictions.}
        (\textbf{A})--(\textbf{C}) Weights assigned to models in three model sets: (A) excluding the worst-fit model, O'~2009, (B) all models, (C) excluding the best-fit, K'~2017.
        (\textbf{D}) RMSE of the posterior mean dose-response prediction for each MMI method and model set.
        Dotted patterning corresponds to (A), no patterning to (B), and dashed patterning to (C).
        The dashed horizontal line is the RMSE of the best-fit model, K'~2017.
        (\textbf{E}) Posterior uncertainty is measured by the mean of the $95\%$ credible interval taken over all EGF levels.
        (\textbf{F}) Relative error of the posterior mean for MMI predictions with increasing model set size.
        All possible combinations of models were tested at each size.
        The dashed blue line shows $5\%$ relative error ($0.05$), and the dashed black line shows the lowest error of any single model, K'~2017.
        (\textbf{G}) Average posterior uncertainty of ERK response for MMI predictions with increasing numbers of models.
        The dashed black line shows the uncertainty of the best model, R' 2015.
        (F),(G) Open circles show outliers.
    }
    \label{fig:synth-mmi-DR-perturb}
\end{figure}

First, single-model perturbations to the model set tended to have small effects on the performance of MMI predictions.
To test this, we first excluded the worst-fit model, O' 2009, (lowest ELPD in Figure~\subpanelref{fig:synth-DR}{C}) and computed the MMI weights (Figure~\subpanelref{fig:synth-mmi-DR-perturb}{A}).
Adding the \textit{bad} model back to the set resulted in little change to the MMI weights (Figure~\subpanelref{fig:synth-mmi-DR-perturb}{B}), and thus, had very little impact on the predictive error (Figure~\subpanelref{fig:synth-mmi-DR-perturb}{D}) and the posterior uncertainty (Figure~\subpanelref{fig:synth-mmi-DR-perturb}{E}) because the additional \textit{bad} model received nearly no weight.
Next, we removed the best-fit model, K' 2017, (highest ELPD in  Figure~\subpanelref{fig:synth-DR}{C}) and recomputed the MMI weights (Figure~\subpanelref{fig:synth-mmi-DR-perturb}{C}).
Without the \textit{best} model, stacking reassigned all weight to L'~2000, the next best model, while pseudo-BMA and BMA redistributed the weight across L'~2000, B'~2007, vK'~2009, S'~2014, and R'~2015.
Accordingly, the predictive error increased very little by approximately $7\%$, $12\%$, and $12\%$ of the error of the worst model, O' 2009, for pseudo-BMA, stacking, and BMA, respectively.
Despite the small increases in error, removing the \textit{best} model had little effect on MMI predictive uncertainty (Figure~\subpanelref{fig:synth-mmi-DR-perturb}{E}).
Therefore, these results show that MMI predictions of ERK activity are robust to changes in the model set because including an additional \textit{bad} model had little effect, and excluding the \textit{best} model led to relatively small changes in uncertainty.

Next, we found that a minimum number of models is needed for accurate MMI, but increasing the number of models beyond that minimum did not vastly improve predictive performance.
To investigate how many models are necessary for effective MMI, we constructed model sets with all possible combinations of the nine ERK models ranging from sets of two to sets of eight models.
In predictions of the EGF-ERK dose-response curve, the relative error of the posterior mean decreased as more models were included for pseudo-BMA and stacking, but not for BMA (Figure~\subpanelref{fig:synth-mmi-DR-perturb}{F}). 
The errors for pseudo-BMA and stacking approached the lowest error of any single model (dashed black line in Figure~\subpanelref{fig:synth-mmi-DR-perturb}{F}), while that for BMA, approached $5\%$ relative error (dashed blue line in Figure~\subpanelref{fig:synth-mmi-DR-perturb}{F}).
Here, we use the relative error of the posterior mean rather than the RMSE because the $5\%$ threshold is easier to interpret. 
Further, for pseudo-BMA and stacking, the fraction of MMI estimates with relative error at or below $5\%$ increased with the number of models in the set (Supplemental Figure~\subpanelref{sup-fig:synth-mmi-DR-modelCombinotorics_quant}{A}).
For pseudo-BMA, approximately $80\%$ of the two-model estimates had relative errors at or below the $5\%$ threshold, and $100\%$ of the eight-model estimates did.
For stacking, those fractions were similarly $80\%$ and $100\%$, for the two-model and eight-model cases, respectively.
However, for BMA, the fraction of estimates with relative errors below the $5\%$ threshold never exceeded $90\%$.
Without any MMI, only $33\%$ of the individual models had relative errors below $5\%$. 
Based on these findings, we conclude that even with just two models, MMI can increase the likelihood of accurately predicting the EGF-ERK dose-response curve beyond what is possible with single models.
However, the improvements were much greater with pseudo-BMA and stacking than with BMA.

The predictive uncertainty decreased with increasing model set size for pseudo-BMA and BMA, but not for stacking (Figure~\subpanelref{fig:synth-mmi-DR-perturb}{H}).
For pseudo-BMA and BMA, the uncertainty approached a level that is below that of the best model (dashed black line in Figure~\subpanelref{fig:synth-mmi-DR-perturb}{H}) as the number of models increased.
With four or more models, $95\%$ of pseudo-BMA and BMA estimates had an uncertainty below the best single model (Supplemental Figure~\subpanelref{sup-fig:synth-mmi-DR-modelCombinotorics_quant}{B}).
However, for stacking, after more than three models are used, we see little uncertainty reduction because the uncertainty is, on average, similar to the prediction made with the best model, K'2017.
These results show that with pseudo-BMA and BMA, four or more models are sufficient to reduce predictive uncertainty, and more models decrease uncertainty only slightly.
Therefore, we conclude that MMI requires a moderate number of models to reduce uncertainty, and that larger model sets do not necessarily lead to greater uncertainty reduction.
Additionally, these results show that even with more model uncertainty (more models in the set), MMI can reliably increase predictive certainty and produce predictions as accurate as the best single models. 

These findings were not limited to dose-response predictions but also extended MMI predictions of dynamic ERK activity.
First, we observed that adding in a poor model or removing the best model had the same effect on trajectory predictions as for dose-response predictions (Supplemental Figure~\subpanelref{sup-fig:synth-mmi-traj-perturb}{A}--\subpanelref{sup-fig:synth-mmi-traj-perturb}{E}).
Next, we found that the error similarly decreases with increasing model count for all methods (Supplemental Figure~\subpanelref{sup-fig:synth-mmi-traj-perturb}{F}), but BMA is the only method that leads to reductions in uncertainty (Supplemental Figure~\subpanelref{sup-fig:synth-mmi-traj-perturb}{D}).
These results show that MMI predictions of dynamic ERK activity are also robust to changes in the model set and that MMI can reduce predictive uncertainty for this prediction type.
In general, we conclude that MMI estimates are robust to additional model uncertainties and reduce predictive uncertainty compared to single models with relatively small model sets.

\subsection{Bayesian multimodel inference predictions are robust to increased uncertainty in the training data} \label{sec:data-quality-quantity}

\begin{figure}[hb!]
    \centering
    \includegraphics[width=\textwidth]{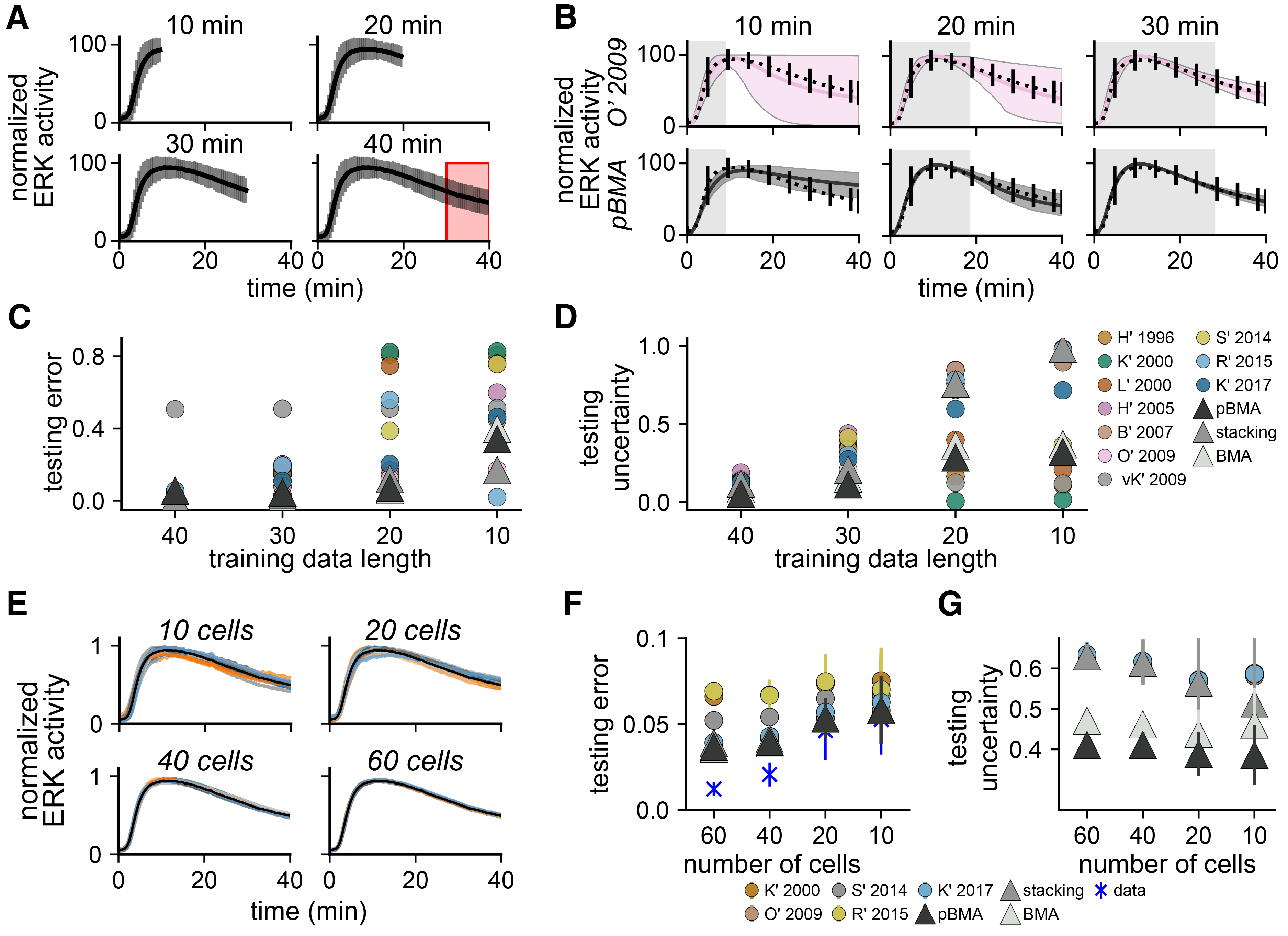}
    \caption{\textbf{Bayesian mulitmodel inference predictions of cytoplasmic ERK activity are robust to uncertainties due to decreasing data length and quality.}
    \textit{Effects of decreasing data length.}
    (\textbf{A}) Shorter training data was constructed by truncating the original 40-minute cytoplasmic ERK trajectories at the 10-, 20-, and 30-minute time points (mean, black trace; and standard deviation, grey bars).
    Predictive performance was assessed by computing errors and uncertainties in the final 10 minutes (red box).
    (\textbf{B}) Posterior predictions from decreased training data for the best model O'~2009 (highest ELPD across all training datasets) and MMI with pseudo-BMA.
    Predictions for additional models and MMI methods are shown in Supplemental Figure~\ref{sup-fig:keyes-data-shorten-SUPP}.
    (\textbf{C}) Predictive error (relative error) for the final 10 minutes ($t=30 \to t=40$ min) of cytoplasmic ERK activity.
    (\textbf{D}) Predictive uncertainty (average width of the $95 \%$ credible interval) for the final 10 minutes of cytoplasmic ERK activity.
    (\textbf{E}) Lower-quality training data was generated by averaging over random subsets of 10, 20, 60, and 60 imaged cells.
    The black trace shows an original average of 76 cells.
    Colored traces show averages of 40 replicate random subsets.
    (\textbf{F}) The predictive error (relative error) of cytoplasmic ERK activity with lower quality data compared to average activity trajectory using all cells.
    (\textbf{G}) Predictive uncertainty of cytoplasmic ERK activity with lower quality data.
    (F),(G) Filled circles indicate average error of 40 replicates for individual models and triangles for MMI predictions.
    Error bars show the standard deviation over replicates.
    Blue markers show the error of the raw training data at each subset size compared to the original full-data mean.}
    \label{fig:keyes-qualityCYTO}
\end{figure}

We have previously observed that data uncertainty can greatly impact predictive uncertainty in intracellular signaling models~\cite{Linden2022-xi}.
To explore how MMI can increase predictive certainty in the face of data uncertainties we used recent experimental observations of cytoplasmic ERK activity from Keyes et al.\ \cite{Keyes2020-ub}.
Using an improved ERK kinase activity reporter called EKAR4,  Keyes et al.\ measured ERK activity at sub-cellular locations, including the cytoplasm and plasma membrane.
With this data, we simulated increasing data uncertainty by decreasing both the quantity and quality of the training data supplied for MMI. 

First, we found that MMI predictions of future ERK activity have lower uncertainty than single-model predictions.
We varied the length of the training data by truncating the original 40-minute long recordings at the 10-, 20-, and 30-minute time points (Figure~\subpanelref{fig:keyes-qualityCYTO}{A}).
The posterior predictions for the best single model, O' 2009, had predictive uncertainty that increased with decreasing data length (top row in Figure~\subpanelref{fig:keyes-qualityCYTO}{B}; additional individual model predictions are shown in Supplemental Figure~\ref{sup-fig:keyes-data-shorten-SUPP}).
However, the pseudo-BMA predictions showed substantially lower predictive uncertainty at all data lengths (bottom row in Figure~\subpanelref{fig:keyes-qualityCYTO}{B}).
Quantitatively, the testing error of MMI predictions is on par with that of the lowest-error models at all training data lengths (Figure~\subpanelref{fig:keyes-qualityCYTO}{C}), and the testing uncertainty of MMI predictions is substantially lower than the best single models (Figure~\subpanelref{fig:keyes-qualityCYTO}{D}).
The testing error and uncertainty measure those quantities in the final 10 minutes ($t=30 \to t=40$) of ERK activity measurements indicated by the red box in Figure~\subpanelref{fig:keyes-qualityCYTO}{A}.
We note that at the 10- and 20-minute data lengths, the L' 2000, K'~2000 and H'~2005 predictions appear more certain than those for MMI, but the predictions for L' 2000, K'~2000 and H'~2005 are extremely erroneous, with almost all simulations predicting maximal activation after 40 minutes (Supplemental Figure~\ref{sup-fig:keyes-data-shorten-SUPP}).
We additionally repeated the same simulations with data from the plasma membrane and observed similar trends as with cytoplasm data (Supplemental Figures~\subpanelref{sup-fig:keyes-qualityPM}{A}--~\subpanelref{sup-fig:keyes-qualityPM}{E}).
Together these results show that MMI increases the certainty and accuracy of predictions of future ERK activity from shorter training data than single models alone.

Second, MMI predictions showed reduced predictive error and uncertainty than single-model predictions from lower-quality data.
We generated lower-quality training datasets by decreasing the number of individual cells that we averaged to create the normalized data.
Specifically, we drew 40 random subsets of 10, 20, 40, and 60 of the 76 original single-cell recordings of cytoplasmic ERK activity and computed the cell-wise average and standard deviation (Figure~\subpanelref{fig:keyes-qualityCYTO}{F}).
Data created with lower cell counts (10 or 20 cells) was of lower quality than data with higher counts (40 or 60 cells) because the lower-count data had greater error compared to the original 76-cell dataset and had more variation in the standard deviation (not shown).
To reduce the computational burden of repeated parameter estimation, we constructed MMI estimates using five of the ten models, K' 2000, O' 2009, S' 2014, R' 2015, and K' 2017, because these models had the shortest inference times (Supplemental Table~\ref{tab:SMC-runtime}).
The predictive testing error, computed as the relative error of model predictions compared to the original 76-cell dataset, shows that MMI predictions are more accurate than most of the individual models (Figure~\subpanelref{fig:keyes-qualityCYTO}{G}).
Additionally, the BMA and pseudo-BMA MMI predictions had lower predictive uncertainty across all subset sizes (Figure~\subpanelref{fig:keyes-qualityCYTO}{H}).
We repeated the same simulations with data from the plasma membrane and similarly found that MMI is robust to lower-quality data (Supplemental Figures~\subpanelref{sup-fig:keyes-qualityPM}{F}--~\subpanelref{sup-fig:keyes-qualityPM}{G}).
Based on these results we conclude the MMI predictions are more robust to uncertainties due to reduced data quality than single-model predictions.
Taken together, both results show that MMI predictions are robust to increases in data uncertainty and, in general, outperform single-model predictions in this context.

\subsection{Sub-cellular location-specific ERK activity depends on Rap1 and ERK negative feedback}\label{sec:ERK-rap1-negFeed}

\begin{figure}[hp!]
    \centering
    \includegraphics[width=\textwidth]{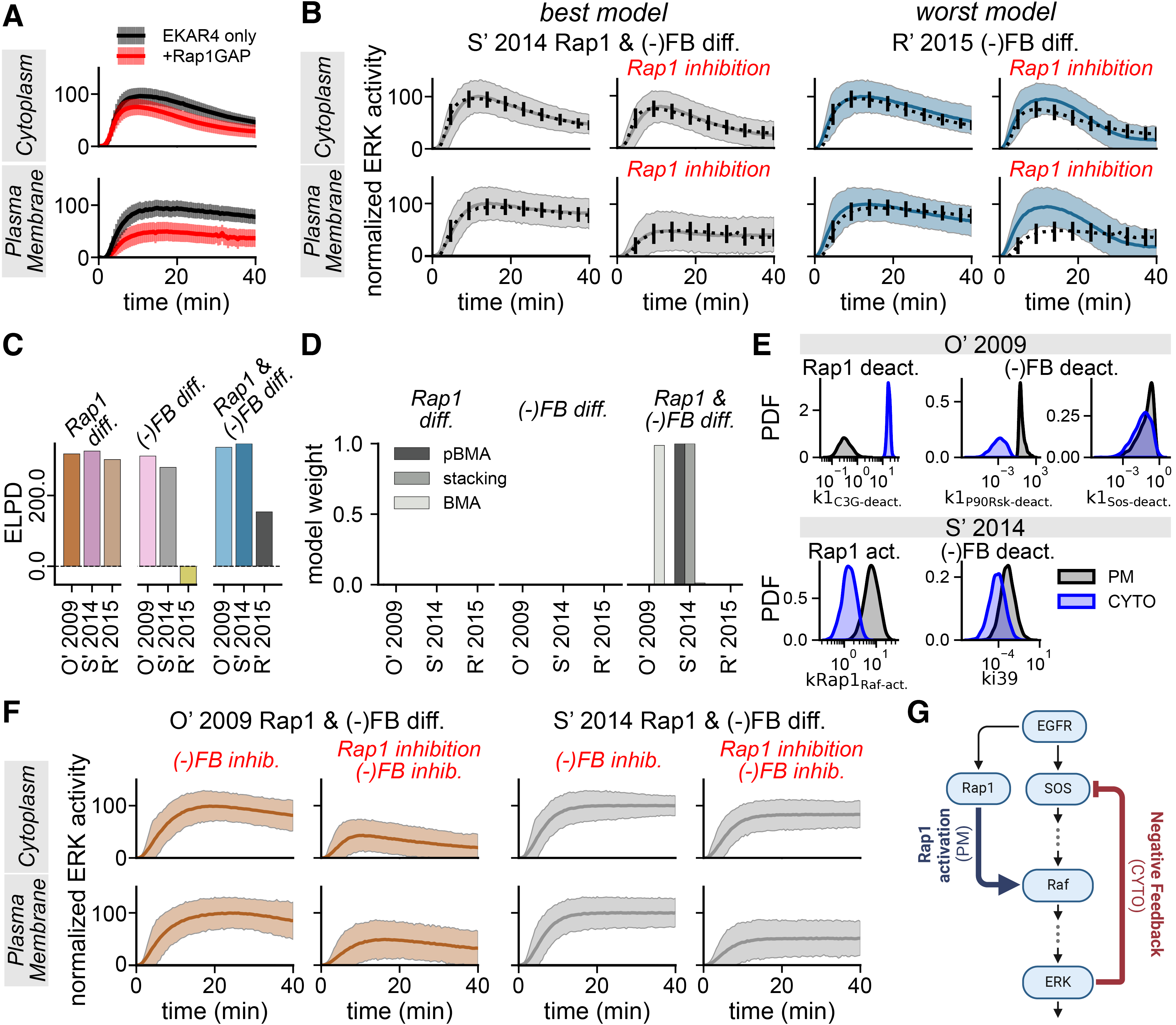}
    \caption{\textbf{Bayesian mulitmodel inference suggests that sub-cellular differences in ERK activity depend on both Rap1 activity and ERK negative feedback.}
    (\textbf{A}) EKAR4 measurements from~\cite{Keyes2020-ub} Figure 3C--3D normalized to the average maximum EKAR4 emission ratio in each location separately.
    Mean and standard deviation for EKAR4-only (black) and EKAR4 with Rap1 inhibition measurements (+Rap1GAP; red).
    (\textbf{B}) Posterior predictions for the best and worst models indicated by ELPD, i.e., S'~2014 with location differences in Rap1 and negative feedback (S'~2014 Rap1 \& (-)FB diff.) and R'~2015 with location differences in negative feedback (R'~2015 (-)FB diff.), respectively.
    Predictions from all models are shown in Supplemental Figure~\ref{sup-fig:keyes-locDiffSUPP}.
    (\textbf{C}) ELPD values for all models.
    (\textbf{D}) MMI weights assigned to all models using pseudo-BMA, stacking, and BMA.
    (\textbf{E}) Marginal posterior densities for the Rap1 activation/deactivation and ERK negative feedback parameters in each location for the O'~2009 and S'~2014 models with Rap1 and negative feedback location differences selected by MMI.
    Cytoplasm densities (CYTO) are in blue, and those for the plasma membrane (PM) are in black.
    All densities are statistically independent between locations ($p< 0.005$ by the Kolmogorov-Smirnov test with a two-sided hypothesis).
    (\textbf{F}) Posterior predictions for selected models with negative feedback inhibition alone ((-)FB inhib.) and simultaneous Rap1 and negative feedback inhibition (Rap1 inhibition + (-)FB inhib.). 
    (\textbf{G}) The proposed mechanism of sub-cellular differences in ERK activity discovered using MMI.
    Rap1 activation is stronger at the plasma membrane, while negative feedback is stronger in the cytoplasm.
    }
    \label{fig:keyes-Rap1-NegFeed}
\end{figure}

The measurements of sub-cellular ERK activity from Keyes et al.\ revealed spatiotemporal differences between cytoplasmic and plasma membrane ERK signaling~\cite{Keyes2020-ub}.
Specifically, the authors observed that ERK activity is more sustained at the plasma membrane compared to the cytoplasm (Figure~\subpanelref{fig:keyes-Rap1-NegFeed}{A}, black traces; data reproduced from figures 1C and 1D of~\cite{Keyes2020-ub}).
Furthermore, Keyes et al.\ showed that plasma membrane ERK activity depended strongly on the non-canonical ERK activator Rap1, but cytoplasmic activity did not (figures 3C and 3D from~\cite{Keyes2020-ub}).
Here, we used MMI to investigate possible mechanisms in the ERK signaling network that drive sub-cellular variations in ERK activity.

Initially, we hypothesized that location-specific probability densities for model parameters can explain location-specific ERK activity observed by Keyes et al.\ \cite{Keyes2020-ub}.
To test this, we estimated ERK signaling model parameters from observations of cytoplasmic and plasma membrane ERK activity independently.
However, we found that the same models can predict both cytoplasmic and plasma membrane ERK activity well when all parameters are allowed to vary independently between the two compartments (Supplemental Figure~\ref{sup-fig:keyes-comp1}).
However, given that most biological processes are tightly regulated~\cite{lavoieERKSignallingMaster2020, kolch2005coordinating}, while mathematically relevant, such sub-cellular location-specific parameter estimates are not physiologically relevant.
Therefore, we hypothesized that Bayesian MMI could predict a more physiologically relevant biochemical mechanism of sub-cellular variability in ERK activity if we restricted the parameters that varied between locations.

To generate models of location-specific ERK signaling for MMI, we focused on estimating parameters related to the noncanonical ERK activator Rap1 and ERK negative feedback.
Specifically, we focus on Rap1 because Keyes et al.\ \cite{Keyes2020-ub} found that plasma membrane ERK activity depends on Rap1 while cytoplasmic activity does not (Figure 3 from~\cite{Keyes2020-ub}.)
Additionally, we included ERK negative feedback because previous findings suggest that the strength of ERK negative feedback can drive differences in sustained versus transient ERK activity~\cite{Orton2009-rd, Brightman2000-ix}.
Thus, we constructed a new set of models that only allowed Rap1 and ERK negative feedback parameters to vary between locations.

To determine if differences in Rap1 kinetics, ERK negative feedback, or both are responsible for the observed location-specific differences, we introduced a Rap1-like activator into the S'2014 and R'2015 models, so that three models, those two and O' 2009, included Rap1.
We assumed that active EGFR activates Rap1, which in turn activates downstream Raf kinase, and that Rap1 activation is not modulated by negative feedback from active ERK~\cite{Orton2009-rd}.
Furthermore, all three models also previously included negative feedback from active ERK to upstream species.
To test specific hypotheses about location-specific mechanisms, we performed nine sets of parameter estimation, restricting the parameters that varied independently between the cytoplasm and plasma membrane.
Specifically, we allowed only the Rap1 parameters (Rap1 diff.), only the ERK negative feedback parameters ((-)FB diff.), or both Rap1 and ERK negative feedback parameters (Rap1 \& (-)FB diff.) to vary independently between locations.
Parameters not associated with Rap1 or ERK negative feedback were assumed to not vary between locations, and a single probability density was estimated for those parameters.
We predicted that MMI would select the models that best reflect the hypotheses that agree with the available data.

Location-specific Rap1 signaling and ERK negative feedback were important in predicting sub-cellular variability in ERK activity.
We fit the models under each hypothesis to EKAR4-only and EKAR4 with Rap1 inhibitor (Rap1GAP) data (Figure~\subpanelref{fig:keyes-Rap1-NegFeed}{A}).
To retain the effects of Rap1 inhibition in each location, we normalized the cell-wise data to the mean of the cell-wise maximum emission ratios and the model predictions to the trajectory-wise maximum with and without Rap1 inhibition.
To simulate Rap1 inhibition, we set the total Rap1 concentration to zero.
Qualitatively, the posterior predictive distributions of the models that allow only ERK negative feedback parameters to vary between locations ((-)FB diff.) predict the data worse, while those that allow Rap1 parameters (Rap1 diff.) predict the data more accurately.
The models that allow both Rap1 and ERK negative feedback parameters to vary, Rap1 \& (-)FB diff., appeared to predict the data best (Figure~\subpanelref{fig:keyes-Rap1-NegFeed}{C} and Supplemental Figure~\ref{sup-fig:keyes-locDiffSUPP}).
The predicted ELPDs show that Rap1 \& (-)FB diff. models do indeed predict the data best, followed by Rap1 diff. and (-)FB diff., respectively (Figure~\subpanelref{fig:keyes-Rap1-NegFeed}{C}).
Furthermore, all three MMI methods only assign weight to models that allow for both Rap1 and negative feedback differences simultaneously (Figure~\subpanelref{fig:keyes-Rap1-NegFeed}{D}).
This suggests that location-specific ERK negative feedback plays a role in location-specific ERK activity, in addition to the Rap1 dependence at the plasma membrane.

Interestingly, MMI predicted that Rap1 activation is stronger at the plasma membrane and ERK negative feedback is stronger in the cytoplasm as indicated by the marginal posterior densities of the associated parameters (Figure~\subpanelref{fig:keyes-Rap1-NegFeed}{E}).
Specifically, the deactivation rate of the Rap1 activator C3G ($k1_{\rm C3G-deact.}$) is lower in the plasma membrane for the O'~2009 model, and the Rap1--Raf activation rate ($ki39$) is higher at the plasma membrane for the S'~2014 model.
Additionally, the deactivation rates of species involved in ERK negative feedback are higher at the plasma membrane than in the cytoplasm, which indicates that negative feedback is stronger in the cytoplasm than in the plasma membrane.
From these results, we conclude that stronger Rap1 activity drives sustained plasma membrane ERK activity, while stronger negative feedback is necessary for transient cytoplasmic ERK activity. 

Indeed, inhibition of ERK negative feedback nearly eliminates location-specific differences in ERK activity (Figure~\subpanelref{fig:keyes-Rap1-NegFeed}{F}).
We simulated ERK negative feedback inhibition by removing corresponding terms from O' 2009 and S' 2014 Rap1 \& (-)FB diff. models. 
Specifically, for O' 2009, we fixed the concentration of p90 RSK, a species only involved in ERK negative feedback, to zero, and for S' 2014, we removed the ERK negative feedback term from the model.
Inhibiting ERK negative feedback at the cytoplasm leads to ERK activation for both models and, therefore, to the loss of location-specific signaling.
However, at the plasma membrane, negative feedback inhibition had little effect on ERK activity.
Thus, these results show that ERK negative feedback, in addition to Rap1 activity, is necessary for location-specific ERK activity.

Together, these findings suggest that cytoplasmic ERK activity depends on ERK negative feedback, while plasma membrane ERK activity depends on Rap1.
Based on the MMI selected hypothesis, we propose a new ERK negative feedback- and Rap1-dependent model for sub-cellular variability in ERK activity (Figure~\subpanelref{fig:keyes-Rap1-NegFeed}{F}).
Here, MMI was critical to selecting the \textit{best} mechanism from the available data without making any \textit{a priori} assumptions that a single hypothesis would be best.

%%%%%%%%%%%%%%%%%%%%%%%%%%%%%%%%%%%%%%%%%%%%%%%%%%%%%%%%%%%%%%%
% DISCUSSION
%%%%%%%%%%%%%%%%%%%%%%%%%%%%%%%%%%%%%%%%%%%%%%%%%%%%%%%%%%%%%%%
\section{Discussion} \label{sec:discussion}

In this work, we introduced a framework for Bayesian multimodel inference in systems biology (Figure~\ref{fig:MMI-overview}).
First, we showed that MMI, specifically pseudo-BMA and BMA, reduces predictive uncertainty by combining multiple models for EGF-ERK dose-response predictions (Figure~\ref{fig:synth-DR}) and predictions of ERK dynamics (Supplemental Figure~\ref{sup-fig:synth-traj}).
In addition to these two examples, we observed uncertainty reduction in MMI predictions of sub-cellular ERK activity (Figure~\ref{fig:keyes-qualityCYTO} and Supplemental Figure~\ref{sup-fig:keyes-comp1}).
Importantly, in all of these examples, MMI estimates retained the predictive accuracy of the best models in the set.
Thus, we suggest that MMI increases the certainty of intracellular signaling predictions because MMI estimates have the same accuracy but lower uncertainty than single-model predictions.

Next, we found that multimodel inference predictions are robust to uncertainties that may be encountered in the modeling process.
First, additional model uncertainty due to adding in a \textit{bad} model or removing the \textit{best} model had only small effects on MMI predictions (Figure~\ref{fig:synth-mmi-DR-perturb} and Supplemental Figure~\ref{sup-fig:synth-mmi-traj-perturb}).
Furthermore, we found that increasing the number of models supplied for MMI improved predictive performance to a point, but beyond that, larger numbers of models led to small additional improvements in predictive performance.
The additional robustness to ``perturbations'' gained from MMI increases predictive certainty because exhaustively including every model for one pathway or \textit{a priori} eliminating \textit{bad models} is often infeasible.
Second, we found that MMI predictions are more robust to additional data uncertainty from reducing data quantity and quality (Figure~\ref{fig:keyes-qualityCYTO} and Supplemental Figure~\ref{sup-fig:keyes-qualityPM}).
Importantly, MMI was able to forecast future ERK activity with greater certainty than individual models from shortened training data which can arise due to the difficulties of live-cell imaging experiments. 
Therefore, we conclude that Bayesian MMI is a powerful approach to combat inevitable modeling and data uncertainties when multiple models are available.

Finally, MMI was critical to discovering that sub-cellular location-specific ERK signaling depends on both Rap1 activity and ERK negative feedback strength.
Specifically, Bayesian MMI heavily weighted models that allowed both Rap1 and ERK negative feedback to vary between compartments, thus suggesting that location-specific differences in Rap1 and ERK negative feedback are important to driving observed differences in ERK activity (Figure~\ref{fig:keyes-Rap1-NegFeed}).
This discovery highlights that MMI can improve certainty in intracellular signaling by selecting \textit{new} models that are most consistent with the available data.
In contrast to standard model selection approaches~\cite{Burnham2002-rl, Kirk2013-pn}, MMI allowed us to avoid the \textit{a priori} assumption that a single model would best capture the observed sub-cellular differences.
From these results, we conclude that MMI increases certainty in systems biology models by improving predictive performance and selecting models that are most consistent with the data.

What is the best multimodel inference method for systems biology problems?
The answer to this question is nuanced and likely application-dependent; however, our findings suggest that pseudo-BMA is a good general method for MMI.
Throughout this work, we found that pseudo-BMA appeared to yield MMI predictions with the lowest error and uncertainty.
Furthermore, pseudo-BMA is widely applicable whenever Bayesian parameter estimation is used due to the generalizability of ELPD approximations such as PSIS-LOO-CV~\cite{Vehtari2016-sb}.
In contrast, stacking most often selects the model with the highest ELPD, and therefore, we conclude that it is best suited for model selection applications.
Further, we do not recommend BMA, because, in addition to its known limitations~\cite{Hoeting1999-hs, Yao2018-gn}, we found that BMA often dangerously reduced predictive uncertainty while increasing predictive error beyond the best models (see, e.g., Figure~\ref{fig:synth-DR}).
While pseudo-BMA proved best in most examples, we found that including multiple MMI methods for model selection applications increased confidence in the selected biological mechanism.
Therefore, based on our findings, we suggest pseudo-BMA as a general approach for Bayesian MMI and only recommend using more than one method for model selection.

Our findings are based on examples of ERK signaling and only utilize one dataset at a time to construct MMI predictions.
While we utilized several different examples of ERK signaling to improve the generalizability of this work, future efforts should similarly compare methods for MMI on additional signaling pathways.
Additionally, while we only used one dataset at a time to construct MMI predictions, incorporating multiple datasets of the same pathway into the MMI framework can potentially improve the confidence in and accuracy of MMI predictions.
Such efforts may potentially require multi-data multimodel inference, which could take inspiration from meta-modeling to weigh datasets in addition to models.

Future technical developments of MMI in systems biology should focus on improving model quality, reducing the computational burden of parameter estimation, and allowing for time-varying MMI weights.
First, our results corroborate previous findings that MMI predictions are only as good as the composition of the set of available models~\cite{Stumpf2020-em}.
Specifically, we found that the predictive accuracy was only as good as the best models in the set (see, e.g., Figures~\ref{fig:synth-DR},~\ref{fig:synth-mmi-DR-perturb} and Supplemental Figures~\ref{sup-fig:synth-traj},~\ref{sup-fig:synth-mmi-traj-perturb}).
Methods to discover new models or expand the model set directly from data~\cite{bruntonDiscoveringGoverningEquations2016a, manganInferringBiologicalNetworks2016}, and to augment models with a discrepancy operator~\cite{morrisonExactReductionGeneralized2021, Kennedy2001-fl} have the potential to improve overall MMI performance by increasing the accuracy of the models.
Second, MMI requires repeating Bayesian parameter estimation for every applicable model, which can introduce a substantial computational burden to a modeling study (see, e.g., inference run times in Supplemental Table~\ref{tab:SMC-runtime}).
Approaches to accelerate Bayesian estimation such as variational inference~\cite{bleiVariationalInferenceReview2017} or Laplace approximation~\cite{Gelman1998-vb} can enable MMI with expensive-to-estimate systems biology models.
Third, this work presents MMI in a framework where the weights on each model do not vary over time. 
However, in experimental datasets that observe large dynamic variations, such as cell-fate transitions, one might expect the underlying mechanism driving the biological mechanism, and thus the most applicable model, to change over time.
In those cases, extensions to the MMI framework, such as dynamic BMA~\cite{rafteryOnlinePredictionModel2010} or sequential data assimilation with multiple models~\cite{Narayan2012-mk} can learn model weights that are time-varying to allow for the MMI estimator to change over time.

With the growing availability of related models of the same intracellular signaling system, Bayesian MMI has the potential to drive new biological discoveries in intracellular signaling.
The rise of systems biology modeling over the past decades and the availability of many models within public databases such as the BioModels Database~\cite{BioModels2020, BioModels2018a} make the availability of multiple models a common reality.
Nevertheless, many systems biology studies only rely on a single model.
From the viewpoint of MMI, working with a single model is comparable to choosing a model at random. Although that model might be the most accurate or \textit{biologically realistic}, the \textit{a priori} choice to use a single model is often made without consideration of the available data.
Throughout this work, we found that Bayesian multimodel inference increases the certainty of predictions when multiple models are available and is thus a powerful approach to identifying new models to describe experimental observations.
Therefore, we conclude that MMI has the potential to improve predictions in systems biology and drive future biological discoveries whenever multiple models of the same system are available.

%%%%%%%%%%%%%%%%%%%%%%%%%%%%%%%%%%%%%%%%%%%%%%%%%%%%%%%%%%%%%%%
% MATERIALS AND METHODS
%%%%%%%%%%%%%%%%%%%%%%%%%%%%%%%%%%%%%%%%%%%%%%%%%%%%%%%%%%%%%%%
\section{Materials and Methods} \label{sec:methods}

\paragraph{Code availability} All code used in this paper is available in our GitHub repository at this link \href{https://github.com/RangamaniLabUCSD/multimodel-inference/}{link}.
In addition to the code, we provide instructions to reproduce each figure in the \verb!README.md! file.

%%%%%%%%%%%%%%%%%%%%%%%%%%%%%%%%%%%%%%%
\subsection{Standardized EGF-inputs} \label{sec:EGF}
To standardize the EGF input across models, we modified all model formulations to include EGF as a state variable.
Then, we set the time derivative of EGF to zero to simulate a sustained EGF stimulus which assumes that modeled reactions do not deplete extracellular EGF pools.
Additionally, we define all EGF stimuli in nanomolar (${\rm nM}$) concentrations.
To convert from concentrations in ${\rm ng}/{\rm mL}$ or ${\rm pg}/{\rm mL}$ to ${\rm nM}$ we assume EGF has a molecular weight of $6{,}048\ {\rm g}/{\rm mol}$~\cite{National-Center-for-Biotechnology-Information2023-fo}.
Additionally, to convert from ${\rm nM}$ to ${\rm molecules}/{\rm cell}$ we assume a cell volume of 1 nanoliter (${\rm nL}$)~\cite{Milo2010-qg}.
Lastly, in models that include additional growth factors, we set the concentrations of the corresponding receptors to zero.
Specifically, in Birtwistle et al.\ 2007~\cite{Birtwistle2007-dw} we fixed concentrations for ${\rm H}$, ${\rm E3}$, and ${\rm E4}$, and in von Kreigsheim et al.\ 2009~\cite{Von_Kriegsheim2009-nq} we fixed the ${\rm NGF}$ and ${\rm NGFR}$ concentrations.

%%%%%%%%%%%%%%%%%%%%%%%%%%%%%%%%%%%%%%%
\subsection{Synthetic data generation and experimental data preprocessing} \label{sec:data-proc}
\paragraph{Synthetic data generation}
We generated two synthetic datasets to analyze MMI.
First, using the H'~1996 model, we recreated the EGF-ERK dose-response curve shown in Figure 2B of~\cite{Huang1996-ki} by varying the input EGF concentration from $0{.}001$ nM to $0.106$ nM over 10 levels and computing the resulting steady-state ERK activity (Figure~\subpanelref{fig:synth-DR}{A}).
Specifically, we normalized each response to the maximum across input levels, and denote this by the \% maximal ERK activity.
Further, we assumed that each measurement is subject to mean-zero Gaussian measurement error with a standard deviation of 0.1.
To account for differences in the total available ERK concentration across models and ensure that model predictions are on the same scale of the data, we computed the \% maximal ERK activity by normalizing each predicted dose-response curve to the maximum ERK activity in that curve.
Second, we again used the H'~1996 model to generate a set of three time-dependent trajectories of ERK activity with EGF concentrations of $0{.}001$ nM, $0{.}005$ nM, and $0{.}106$ nM (Supplemental Figure~\subpanelref{sup-fig:synth-traj}{A}).
Data and model predictions are normalized to the maximum value of trajectory with simulated with [EGF]$ = 0{.}106$ nM.
We again assumed that data was subject to mean-zero Gaussian measurement error with a standard deviation of 0.1 and let each trajectory be 30 minutes long to ensure that a steady state was approximately reached at each stimulus level.

\paragraph{EKAR4 data preprocessing}
Keyes et al.\ report all EKAR measurements as the YFP/CFP emission ratio, $R(t)$~\cite{Keyes2020-ub}.
In Figure~\ref{fig:keyes-Rap1-NegFeed}, we define the normalized EKAR4 data as $\tilde{R}(t) \coloneqq R(t)/\overline{R_{\rm max}}$,
where $R(t)$ is the EKAR4 emission ratio, and $\overline{R_{\rm max}}$ is the mean (across all cells with and without Rap1 inhibition) of the maximum (in time) emission ratio.
This choice of normalization retains both location-specific differences in ERK signaling and the effects of Rap1 inhibition.
Further, in Supplemental Figure~\ref{sup-fig:keyes-comp1}, we normalize each cell-wise trajectory by removing the cell-wise minimum and dividing by the difference in the cell-wise maximum and minimum. 
All original data are reported as measurements over individual cells, thus to yield a single dataset with an uncertainty estimate, we computed the average and standard deviation over all cells (e.g., black and red curves in Figure~\subpanelref{fig:keyes-Rap1-NegFeed}{A}).
We found that normalization introduced a strong uncertainty in Rap1 inhibition data, so in constructing a likelihood model, we took the standard deviation of the data to be one-half of that computed directly (shown as gray and red bars in Figure~\subpanelref{fig:keyes-Rap1-NegFeed}{A}).
To ensure that the data and model predictions were on the same scale, we similarly normalized model predictions of active ERK to the trajectory-wise maximum.

\subsection{Bayesian parameter estimation and sequential Monte Carlo sampling} \label{sec:bayesian-calib-ODE-METHODS}

We provide a brief overview of Bayesian parameter estimation in this section.
For more details in the context of systems biology, see~\cite{Linden2022-xi, Geris2016-xv}, and more theory in general, see~\cite{Smith2013-lw, Gelman2014-jj}.

Systems biology models use systems of \textit{ordinary differential equations} (ODEs) to describe the rates of change in the concentration of included biochemical species.
For a particular signaling pathway, we often have a set of $K$ models, $\fM_K = \left\{\cM_1,\ldots,\cM_K\right\}$, that vary in their representation of the system.
Each model connects the dynamics of the state variables $\bx_k(t) \in \real_+^{n_k},$ ($\real_+ = [0, \infty)$ are the nonnegative real numbers) which correspond to the concentration of biochemical species, to observations $\hat{\by}_k(t) \in \real_+^{m_k}$ using the system of equations
\begin{align}
    \frac{\rd \bx_k(t)}{\rd t} &= f_k(\bx_k(t); \btheta_k), \label{eq:diffyq-ode} \\
    \hat{\by}_k(t) &= h_k(\bx_k(t); \btheta_k) + \boldsymbol{\eta}(t) \quad \boldsymbol{\eta}(t) \sim \cN(\mathbf{0}, \bGamma), \label{eq:meas-func-ode}
\end{align}
where $f_k(\cdot; \cdot): \real_+^{n_k} \times \real_+^{p_k} \to \real_+^{n_k}$, and
$h_k(\cdot; \cdot): \real_+^{n_k} \times \real_+^{p_k} \to \real_+^{m_k}$.
The model parameters $\btheta_k \in \Theta_k \subseteq \real_+^{p_k}$ control model predictions and include quantities such as reaction rates and equilibrium coefficients.
We additionally assume that a Gaussian measurement noise process $\boldsymbol{\eta}(t) \in \mathbb{R}^{m_k}$ with covariance matrix $\bGamma \in \mathbb{R}^{m_k \times m_k}$ reflects uncertainty in the measurements.
Beyond predictions of the observations, we assume that each model predicts a biologically relevant quantity of interest (QoI) $q(t) \in \mathbb{R}$.
The QoI is predicted by a function of the internal states and parameters, $\hat{q}_k = g_k(\bx_k(t), \btheta_k)$, where $\hat{q}_k$ is the QoI prediction with model $\cM_k$, and $g_k(\cdot, \cdot): \real_+^{n_k} \times \real_+^{p_k} \to \real $.

Bayesian parameter estimation learns a probability density for the parameters of each model conditioned on the training data ${\rm p}(\btheta_k | \cD_{\rm train}, \cM_k)$~\cite{Gelman2014-jj, Smith2013-lw}.
The training data $\cD_{\rm train} = \{\by^1, \ldots, \by^{D_{\rm train}}\}$ consists of $D_{\rm train}$ noisy experimental observations, and can correspond to time points $t^i$ in dynamic responses $\by^i = \by(t^i)$ or to input stimuli $u^i$ in dose-response curves $\by^i = \by(u^i)$.
Bayesian estimation applies Bayes' rule
\begin{equation}
        \underbrace{{\rm p}(\btheta_k | \cD_{\rm train}, \cM_k)}_{\rm posterior} \propto \underbrace{{\rm p}(\btheta_k | \cM_k)}_{\rm prior} \underbrace{{\rm p}(\cD_{\rm train} | \btheta_k)}_{\rm likelihood}, \label{eq:bayes-rule}
\end{equation}
which relates the \textit{posterior probability density} of the model parameters, ${\rm p}(\btheta_k | \cD_{\rm train}, \cM_k)$, to the product of the \textit{prior density} $ {\rm p}(\btheta_k | \cM_k)$ and the \textit{likelihood function} ${\rm p}(\cD_{\rm train} | \btheta_k)$.
The posterior density is the probability density of the model parameters given the available training data and the model.
The prior density encodes assumptions about the parameters before considering training data.
The likelihood function measures the probability that model predictions match the training data and is a function of the model parameters.
In most systems biology problems, we do not have a closed-form equation to evaluate the posterior density directly, so we instead must rely on methods such as \textit{Markov chain Monte Carlo} (MCMC)~\cite{Gelman2014-jj, Smith2013-lw, Linden2022-xi} or \textit{sequential Monte Carlo} (SMC)~\cite{chingTransitionalMarkovChain2007,minsonBayesianInversionFinite2013} to characterize the posterior through the $S$ samples drawn from it, $\{\btheta_k^1, \dots, \btheta_k^S\} \sim {\rm p}(\btheta_k | \cD_{\rm train}, \cM_k)$.

This work uses log-normal prior densities for all unknown model parameters.
To let the data inform the estimation procedure, we choose the mean of each prior to being centered on the logarithm of the nominal values for the unknown parameters $\btheta_k^{\rm nominal}$ and let the standard deviation be suitably large such that $95\%$ of the probability mass of the prior is in the range $[10^{-2} \times \btheta_k^{\rm nominal}, 10^2 \times \btheta_k^{\rm nominal}]$.
One can show that this corresponds to setting the prior standard deviation to $2.350$.
Substantial empirical evidence suggests that wide, weakly informative priors such as the lognormal prior with wide variance greatly enable sampling of the posterior density compared to completely uninformative priors such as the uniform prior~\cite{Gelman2014-jj}.
Next, the likelihood function is a Gaussian density because we assume a Gaussian measurement noise in Eq.~\eqref{eq:meas-func-ode}~\cite{Linden2022-xi}.
We use the PyMC probabilistic programming library in Python to efficiently build statistical models and enable posterior sampling~\cite{pymc2023}.

We use sequential Monte Carlo to sample from the posterior density because it estimates the marginal likelihood without additional computation~\cite{chingTransitionalMarkovChain2007,minsonBayesianInversionFinite2013}.
SMC is a particle-based sampler that sequentially adapts the particles from prior samples to posterior samples using a tempering scheme.
The particles are weighted and mutated at each sampler stage with an importance sampling step.
At the final stage, the weights assigned to the particles correspond to the marginal likelihood and can be averaged to provide a marginal likelihood estimate.
We use the implementation provided by the PyMC python package with the independent Metropolis-Hastings transition kernel.
Additionally, we set the \verb!correlation_threshold! parameter to 0.01 and the \verb!threshold! parameter to 0.85.
Unless otherwise noted, we run four independent SMC chains with at least 500 posterior samples per chain.

\subsection{Forward uncertainty propagation with ensemble simulation} \label{sec:for-UQ}
Given $S$ samples from the posterior density, we perform ensemble simulations to propagate uncertainty forward to model predictions.
Specifically, we solve the ODE model with each of the $S$ posterior samples to generate a set of ODE solutions for resulting predictive densities.
There are three important predictive densities that we use to assess predictive uncertainty. 
First, the \textit{posterior push-forward density of the QoI} (also called the predictive density of the QoI) is defined as 
\begin{equation}
    {\rm p}(\hat{q}_k|\cD_{\rm train}, \cM_k) \coloneqq \int_{\Theta_k} g_k(\bx_k(t), \btheta_k) {\rm p}(\btheta_k|\cD_{\rm train}, \cM_K) \rd \btheta_k
\end{equation}
and directly propagates parametric uncertainty to the QoI.
To sample from the posterior push-forward density of the QoI, we evaluate the QoI function $g_k(\cdot, \cdot)$ at each sample, yielding a set of QoI samples $\{\hat{q}_k^1, \dots, \hat{q}_k^S\} \sim {\rm p}(\hat{q}_k|\cD_{\rm train}, \cM_k)$.
Next, the \textit{posterior push-forward density of the measurements}, 
\begin{equation} \label{eq:post-pushF}
    {\rm p}(\hat{\by}(t)_k|\cD_{\rm train}, \cM_k) \coloneqq \int_{\Theta_k} h_k(\bx_k(t)) {\rm p}(\btheta_k|\cD_{\rm train}, \cM_K) \rd \btheta_k,
\end{equation}
similarly propagates parametric uncertainty to the measurements.
We evaluate the measurement function at each ODE solution to sample from the posterior push-forward density of the measurements.
Lastly, the \textit{posterior predictive density} is defined as
\begin{equation} \label{eq:post-pred}
    {\rm p}(\tilde{\cD}_{\rm train}|\cD_{\rm train}, \cM_k) \coloneqq \int_{\Theta_k} {\rm p}(\cD_{\rm train}|\btheta_k, \cM_k) {\rm p}(\btheta_k|\cD_{\rm train}, \cM_K) \rd \btheta_k,
\end{equation}
and accounts for both parametric and data uncertainty.
To obtain posterior predictive samples, we sample the posterior push-forward density of the measurements and subsequently add independent samples from the measurement noise process defined in Eq.~\eqref{eq:diffyq-ode}.
We refer the reader to~\cite{Martin2021-ly}~and~\cite{Gelman2014-jj} for more details on predictive densities and Bayesian model analysis in general.

In this work, we use both the posterior push forward of the QoI and the posterior predictive density where appropriate.
In Figures~\ref{fig:synth-DR},~\ref{fig:synth-mmi-DR-perturb},~and~\ref{fig:keyes-Rap1-NegFeed} and Supplemental Figures~\ref{sup-fig:synth-mmi-traj-perturb}~and~\ref{sup-fig:keyes-comp1}, we use the posterior predictive density because the QoI function is equivalent to the measurement function.
Alternatively, in Figure~\ref{fig:keyes-qualityCYTO} and Supplemental Figure~\ref{sup-fig:keyes-qualityPM}, we use the posterior push forward of the QoI because the QoI function is not the measurement function.

\subsection{Bayesian model averaging} \label{sec:BMA}
Bayesian model averaging weighs each model with the model probability $w_k^{\rm BMA} = {\rm p}(\cM_k |\cD_{\rm train})$~\cite{Hoeting1999-hs}.
Notably, the model probability is the realization of a discrete probability mass function over the set of models.
The model probability is computed by applying Bayes' rule a second time (the first time is for model parameter estimation) at the model level, that is
\begin{equation}
    {\rm p}(\cM_k |\cD_{\rm train}) = \frac{{\rm p}(\cD_{\rm train}|\cM_k){\rm p}(\cM_k)}{\sum_{l=1}^K {\rm p}(\cD_{\rm train}|\cM_l){\rm p}(\cM_l)}, \label{eq:BMA}
\end{equation}
where ${\rm p}(\cM_k)$ is the prior model probability and ${\rm p}(\cD_{\rm train}|\cM_k)$ is the marginal likelihood.
Importantly, the marginal likelihood is included in Eq.~\eqref{eq:BMA} because it depends on the model, whereas we exclude the denominator in parameter estimation and replace equality with proportionality in Eq.~\eqref{eq:bayes-rule}.
The \textit{marginal likelihood},
\begin{equation}
    {\rm p}(\cD|\cM_k) = \int_{\Theta_k} {\rm p}(\cD|\btheta _k, \cM_k){\rm p}(\btheta_k|\cM_k)\rd \btheta_k, \label{eq:BMA-marginal_like}
\end{equation}
quantifies the probability of observing the data under model $\cM_k$.
Direct computation of Eq.~\eqref{eq:BMA-marginal_like} can become intractable because the integral is often over a high-dimension space and thus requires approximations by Monte Carlo integration, Bridge Sampling, or sequential Monte Carlo~\cite{Kass1995-xl, Gronau2017-tz, Hoeting1999-hs, George2014-mx}.
In this work, we use the PyMC implementation of sequential Monte Carlo, which estimates the log marginal likelihood alongside posterior samples~\cite{pymc2023,chingTransitionalMarkovChain2007,minsonBayesianInversionFinite2013}.
PyMC returns a log marginal likelihood estimate for each independent chain, so we take the average over all chains to generate an estimate for that model.
Additionally, we assume that each model is equally probable \textit{a priori} with prior model probability ${\rm p}(\cM_k) = 1/K$.
To avoid numerical overflow in evaluating Eq.~\eqref{eq:BMA}, which involves computing the sum of the exponent of log marginal likelihoods, we employ the log-sum exponential trick.
That is, given log marginal likelihoods, Eq.~\eqref{eq:BMA} can be written abstractly as
\begin{equation}
     {\rm p}(\cM_k |\cD_{\rm train}) = \frac{\exp(x_k)}{\sum_{l=1}^K \exp(x_l)}, \label{eq:ml-abstract}
\end{equation}
where $x_k = \log \left({\rm p}(\cD_{\rm train}|\cM_k){\rm p}(\cM_k)\right)$ can be quite large and thus Eq.~\eqref{eq:ml-abstract} is susceptible to overflow errors.
We can rearrange Eq.~\eqref{eq:ml-abstract} to
\begin{equation*}
     {\rm p}(\cM_i |\cD_{\rm train}) = \exp \left[x_i - \log \sum_{l=1}^K \exp(x_l) \right], \label{eq:new}
\end{equation*}
where we use the log-sum-exponential trick 
\begin{equation*}
    \log \sum_{l=1}^K \exp(x_l) = c + \log \sum_{l=1}^K \exp(x_l - c), \label{eq:LSE}
\end{equation*}
with $c=\max[x_1, \dots, x_K]$ for numerical stability.

\subsection{Pseudo-Bayesian model averaging} \label{sec:pseudoBMA}
%%%%%%%%%%%%%%%%%%%%%%%%%%%%%%%%%%%%%%%

Pseudo-Bayesian model averaging weighs models based on the expected predictive performance of future data measured with the expected log pointwise predictive density (ELPD).
Similar to Akaike-type weighting, which uses the Akaike information criterion instead of the ELPD~\cite{Burnham2002-rl}, we define the scalar pseudo-BMA weights as
\begin{equation} \label{eq:pbma-weights}
    w_k^{\rm pBMA} \coloneqq \frac{\exp(\widehat{\rm ELPD}_{\cM_k})}{\sum_{i=1}^K \exp(\widehat{\rm ELPD}_{\cM_k})}, 
\end{equation}
given an estimate of the ELPD for each model $\widehat{\rm ELPD}_{\cM_k}$.
Further, we stabilize Eq.~\eqref{eq:pbma-weights} with the log-sum-exponential trick as in the previous section.
To account for uncertainty in the ELPD estimates, we use the Bayesian bootstrap as was done in~\cite{Rubin1981-ta, Yao2018-gn}.

We adapt the definitions from~\cite{Vehtari2016-sb} to introduce approximations to the ELPD that can be used for pseudo-BMA.
First, we let the training data $\cD_{\rm train} = \{\by^1, \ldots, \by^{D_{\rm train}}\}$ consist of statistically independent data points.
It follows that ${\rm p}(\cD_{\rm train}|\btheta_k, \cM_k) = \prod_{i=1}^{D_{\rm train}} {\rm p}(\by^i|\btheta_k, \cM_k)$ due to the independence of the data.
The \textit{expected log pointwise predictive density} of model $\cM_k$ is defined as
\begin{equation*} \label{eq:elpd}
    {\rm ELPD}_{\cM_k} \coloneqq \sum_{i=1}^{D_{\rm train}} \int \log \ {\rm p}(\tilde{\by}_i|\cD_{\rm train}, \cM_k) {\rm p}_{\rm true}(\tilde{\by}_i) \rd \tilde{\by}_i,
\end{equation*}
and quantifies the expected predictive performance of the model compared to the true data generating distribution ${\rm p}_{\rm true}(\tilde{\by}_i)$.
The \textit{log posterior predictive density} for model $\cM_k$ and new data-point $\tilde{\by}_i$ is defined as
\begin{equation*}
    \log \ {\rm p}(\tilde{\by}_i|\cD_{\rm train}, \cM_k) \coloneqq \log \int_{\Theta_k}{\rm p}(\tilde{\by}_i|\btheta_k, \cM_k){\rm p}(\btheta_k|\cD_{\rm train}, \cM_k) \rd \btheta_k
\end{equation*}
 and measures the log probability of observing the new data point.
 Given $S$ posterior samples $\hat{\btheta}_k^s$, the log posterior predictive density can be approximated with
\begin{equation} \label{eq:post-pred-MC}
    \log \ {\rm p}(\tilde{\by}_i|\cD_{\rm train}, \cM_k) \approx \log \frac{1}{S}\sum_{i=1}^S {\rm p}(\tilde{\by}_i|\hat{\btheta}_k^s, \cM_k).
\end{equation}
In general, we do not know the true probability of the data ${\rm p}_{\rm true}(\tilde{\by}_i)$, so we must rely on approximations to the ELPD such as the leave-one-out cross-validation estimator (LOO-CV)~\cite{Vehtari2016-sb}, which we use in this work, or alternatively, the widely applicable information criterion (WAIC)~\cite{watanabeAsymptoticEquivalenceBayes2010}.

The simplest approximation to the ELPD, called the \textit{sample-approximated log pointwise predictive density} (LPD), is simply the sum of Eq.~\eqref{eq:post-pred-MC} over all data points,
\begin{equation*} \label{eq:comput-lppd}
    \widehat{\rm LPD}_{\cM_k} = \sum_{i=1}^{D_{\rm train}} \log \frac{1}{S}\sum_{s=1}^S {\rm p}(\by_i|\hat{\btheta}_k^s, \cM_k);
\end{equation*}
however, the LPD is known to overestimate the ELPD~\cite{Vehtari2016-sb}.
Thus, we can use LOO-CV to approximate the ELPD as
\begin{equation*}
        \widehat{{\rm ELPD}}_k^{\rm LOO} \coloneqq \sum_{i=1}^n \log {\rm p}(\by_i | \cD^{-i}_{\rm train}, \cM_k), \label{eq:elpd-loo}
\end{equation*}
where the \textit{leave-one-out predictive density} is defined as
\begin{equation}
    {\rm p}(\by^i | \cD^{-i}_{\rm train}, \cM_k) \coloneqq \int {\rm p}(\by^i | \btheta_k, \cM_k) {\rm p}(\btheta_k|\cD^{-i}_{\rm train})\rd \btheta_k. \label{eq:loo-pred}
\end{equation}
Here,  $\cD^{-i}_{\rm train} \coloneqq \{\by^1, \dots, \by^{i-1}, \by^{i+1}, \dots, \by^{D_{\rm train}}\}$ indicates all of the training data $\cD_{\rm train}$ excluding point $\by^i$.
We refer the reader to Vehtari et al.\ \cite{Vehtari2016-sb} for further details and motivation.
This work uses the LOO-CV-based approximation to the ELPD because both approximations provided similar weights for each model (results not shown here).
However, direct computation of Eq.~\eqref{eq:loo-pred} can be prohibitively computationally expensive because it requires repeating parameter estimation for each held-out data point.
To enable efficient computation of the LOO-CV ELPD estimator, we use Pareto smoothed importance sampling~\cite{Vehtari2022-tj, Vehtari2016-sb}, the details of which we omit for brevity.
We denote the Pareto smoothed importance sampling leave-one-out cross-validation estimator of the ELPD as PSIS-LOO-CV.

In this work, we use the ArviZ Python library~\cite{arviz_2019} to compute the PSIS-LOO-CV and the resulting model weights.
Given a dictionary of inference data objects with log-likelihood samples for each model, we use the \verb!Arviz.compare()! function with the \verb!ic! parameter set to \verb!loo! and the \verb!method! parameter set to \verb!BB-pseudo-BMA!.
We use default settings for all computations.
To ensure that PSIS-LOO-CV accurately estimates the LOO-CV estimator, we compared PSIS-LOO-CV to direct leave-one-out cross-validation (LOO-CV) for the S'~2014 model using synthetic dose-response data (Figure~\subpanelref{fig:synth-DR}{A}).
Notably, the ELPD predictions for PSIS-LOO-CV and LOO-CV are within 5\% percent, predicting $-5.69$ and $-5.43$, respectively.

%%%%%%%%%%%%%%%%%%%%%%%%%%%%%%%%%%%%%%%
%% STACKING
\subsection{Stacking of predictive densities} \label{sec:stacking}
Stacking of predictive densities~\cite{Yao2018-gn} aims to assign scalar weights $w_k^{\rm stack}$ to each model such that the new estimator is optimal according to a specified optimality criterion.
The stacking optimization problem
\begin{equation}
    \underset{\{w_k^{\rm stack}\}}{\max} \ S\left(\sum_{k=1}^K w_k^{\rm stack} {\rm p}(\tilde{\cD} | \cD_{\rm train}, \cM_k), {\rm p}_{\rm true}(\tilde{\cD})\right) \label{eq:stacking}
\end{equation}
aims to maximize a score, $S(\cdot, \cdot): \mathcal{P} \times \Omega \to \real$, where $\mathcal{P}$ is a probabilistic forecast and $\Omega$ is the sample space on which the true predictive density is defined (see~\cite{Yao2018-gn} for a more detailed definition).
In this case, the score is computed between the consensus predictive density evaluated at new data $\tilde{\cD}$ and the true predictive density.
However, the true predictive density is unknown, so Yao et al.\ \cite{Yao2018-gn} suggest using a logarithmic scoring rule and replacing the predictive density evaluated at new data with the LOO estimator.
The approximate stacking optimization problem becomes
\begin{equation}
    \underset{\{w_k^{\rm stack}\}}{\max} \ \frac{1}{n} \sum_{i=1}^n \log \sum_{k=1}^K w_k^{\rm stack} {\rm p}(\by_i| \cD^{-i}_{\rm train}, \cM_k),\label{eq:stacking-log-LOO}
\end{equation}
where the LOO-CV predictive density ${\rm p}(\by_i| \cD^{-i}_{\rm train}, \cM_k)$ can again be estimated with PSIS-LOO-CV as in the previous section.

We again use the ArviZ Python package~\cite{arviz_2019} to compute the stacking weights in this work.
Specifically, we use the \verb!Arviz.compare()! function with the \verb!ic! parameter set to \verb!loo! and the \verb!method! parameter set to \verb!stacking!.
All other parameters are set to the defaults.

%%%%%%%%%%%%%%%%%%%%%%%%%%%%%%%%%%%%%%%%
% ODE Simulation
\subsection{Numerical solution of ODEs and steady-state simulation} \label{sec:ode-sim}

We solve all systems of ordinary differential equations with the Kvaerno 4/5 implicit Runge-Kutta method~\cite{kvaerno2004singly} that is implemented in the Diffrax Python package~\cite{kidger2021on}.
Unless otherwise stated, we set the maximum number of solver steps, \verb!max_steps!, to $6 \times 10^6$.
Additionally, we use a PID controller-based method for adaptive time stepping~\cite{soderlind2003digital, soderlind2002automatic, hairer2002solving-ii} with tolerances \verb!atol=1e-6! and \verb!rtol=1e-6! for all problems unless otherwise noted in Supplemental Table~\ref{tab:ODE-hyperparam}.
We run the solver from the initial condition to the desired time point to obtain the entire solution over time.
However, we choose the following methods to balance accuracy and computational efficiency to obtain each model's steady-state solution.
The first approach runs the ODE solver until the solution and time derivative satisfy $||\rd \bx(t)/\rd t||_{2} \leq {\rm atol} + {\rm rtol}||\bx(t)||_{2}$, where $||\cdot||_2$ is the standard vector two-norm.
To evaluate this criterion, we use a Diffrax \verb!SteadyStateEvent! with tolerances \verb!atol=1e-5! and \verb!rtol=1e-6! for all problems unless stated otherwise in Supplemental Table~\ref{tab:ODE-hyperparam}.
The second approach uses Newton's method to directly solve $\rd \mathbf{x}(t)/\rd t = 0$.
We use the standard Newton method implemented in the Optimistix Python library~\cite{optimistix2024} with \verb!max_steps = 100! and tolerances \verb!atol=1e-10! and \verb!rtol=1e-10!.
To mitigate convergence issues, we initialize the Newton solver at a point along the solution of the ODE by running the previous ODE-based steady-state method with crude tolerances \verb!atol=1e-5! and \verb!rtol=1e-5! for all problems unless stated otherwise in Supplemental Table~\ref{tab:ODE-hyperparam}.

%%%%%%%%%%%%%%%%%%%%%%%%%%%%%%%%%%%%%%%%
% identifiability analysis
\subsection{Structural Identifiability Analysis}\label{sec:struct-id}

We perform an \textit{a priori} local structural identifiability analysis to reduce the unknown model parameters that have unique values given the observables.
We refer the reader to~\cite{Wieland2021-tz, Dong2023-ak} for background and mathematical details on structural identifiability analysis.
We perform a local analysis, as opposed to a global analysis as suggested in our previous work~\cite{Linden2022-xi}, because computing global identifiability proved computationally intractable for larger models such as H'~2005 and B'~2007.
Specifically, we used the structural identifiability analysis method from~\cite{Dong2023-ak} that is implemented in the StructuralIdentifiability.jl package in the Julia programming language.
We use default settings for the \verb!assess_local_identifiability()! function, and set the probability of correctness to $p=0.99$ for all models except Birtwistle et al.\ 2007~\cite{Birtwistle2007-dw}, for which we take $p=0.95$ to improve computational efficiency.
Additionally, due to further software limitations, we exclude any parameters in exponents from the identifiability analysis.
If these parameters are integers, we fix them to their nominal value, otherwise we set them to 1.0.
Specifically, for Hornberg et al.\ 2005~\cite{Hornberg2005-fs} we fix $n$ to $1.0$, and for von Kreigsheim et al.\ 2009~\cite{Von_Kriegsheim2009-nq} we fix $k_{57}$, $k_{61}$, $k_{64}$, $k_{66}$, $k_{70}$, and $k_{72}$ to $1.0$.

\subsection{Global Sensitivity Analysis} \label{sec:gsa}

We use the Morris screening method for global sensitivity analysis~\cite{Morris1991-zd, Campolongo2007-ak}.
First, we assume that all parameters vary independently and can take on values in the range $[10^{-2} \times \theta^{\rm nominal}, 10^2 \times \theta^{\rm nominal}]$ (nominal values are listed in Supplemental Materials).
We then use the steady-state activated ERK concentration as the quantity of interest for models that predict sustained ERK activation at the nominal parameter values, i.e., H'~1996, O'~2009, S'~2014~, R'~2015, and K'~2017, or the maximal activated ERK concentration as the QoI for models that predict a transient activation, i.e., K'~2000, L'~2000, H'~2005, B'~2007, and vK'~2009.
Additionally, we assume that the EGF concentration is fixed to 0.1 nM for all models.
To perform the Morris screening, we use the SALib Python package~\cite{Iwanaga2022, Herman2017} with the method of Morris sampler (\verb!SALib.sample.morris.sample()!) and analysis  (\verb!SALib.sample.moris.analyze()!) functions with default settings.
For sampling, we draw 256 samples per parameter direction for all models except H'~2005 and B'~2007 for which we use 30 and 10 samples, respectively.
Morris screening provides two measures of sensitivity, the mean of the distribution of the absolute value of elementary effects $\mu^*$ and the standard deviation of the distribution of elementary effects $\sigma$~\cite{Campolongo2007-ak}.
We refer the reader to~\cite{Campolongo2007-ak, saltelliGlobalSensitivityAnalysis2007} for more details.
We assume that parameters are influential to ERK activation when $\mu_i^*/\max\{\mu_i^*\} > 0.1$ or $\sigma_i/\max\{\sigma_i^*\} > 0.1$.

\subsection{Error and uncertainty metrics}
We measure predictive error with either the root mean square error (RMSE) or the relative error.
Given a prediction $\hat{\by}$ and a reference (ground truth) $\by$,
the \textit{RMSE} is defined as,
\begin{equation*}
    {\rm RMSE} \coloneqq \frac{\sqrt{\sum_{i=1}^N(\hat{y}_i - y_i)^2}}{N},
\end{equation*}
where $\by \in \real^N$, and $y_i$ is the $i$th element of $\by$. We additionally define the \textit{relative error} as
\begin{equation*}
    {\rm Relative \ Error} \coloneqq \frac{||\hat{\by} - \by||_2}{||\by||_2},
\end{equation*}
where the norm is the standard vector two-norm, $||\ba||_2 = \sqrt{\sum_{i=1}^N a_i^2}$.

We measure predictive uncertainty using the average width of the $95\%$ credible interval.
Given a set of predictive samples prediction, ${\hat{\by}^1, \dots, \hat{\by}^S}$, we define the \textit{$95\%$ credible interval} as the interval between the $2.5$th and $97.5$th percentiles,
\begin{equation*}
    95\% \ {\rm credible \ interval} \coloneqq [P_{2.5}, P_{97.5}],
\end{equation*}
where $P_{i}$ denotes the element-wise $i$th percentile of the set of samples.
That is, each element of $P_{i}$ is the percentile of the corresponding set of elements from the set of vector-valued predictive samples.
We define the average width of the $95\%$ credible interval as the mean over all elements of the difference between $P_{97.5}$ and $P_{2.5}$.

\subsection{Statistical comparison of probability densities}
We compared probability densities using either the Kolmogorov-Smirnov test (K--S Test) or the Mann-Whitney U-test with two-sided hypotheses.
All statistical comparisons were performed using the Scipy Python library.

%%%%%%%%%%%%%%%%%%%%%%%%%%%%%%%%%%%%%%%%
% BIBLIOGRAPHY
%%%%%%%%%%%%%%%%%%%%%%%%%%%%%%%%%%%%%%%
% \printbibliography
\bibliographystyle{IEEEtran}
\bibliography{refs}

%%%%%%%%%%%%%%%%%%%%%%%%%%%%%%%%%%%%%%%%
% SUPPLEMENTAL MATERIALS
%%%%%%%%%%%%%%%%%%%%%%%%%%%%%%%%%%%%%%%
\newpage
\begin{appendix}
    % Change figure labeling and renew counter
    \renewcommand{\thefigure}{S\arabic{figure}}
    \setcounter{figure}{0}
   \section{Supplemental Figures} \label{sec:supp-figs}

%%%%%%%%%%%%%%%%%%%%%%%%%%%%%%%%%%%%%%%%%%%%%%%%%%%%%%%%%%%%%%%%%%%
% Supp fig -- GSA -- goes with results section 2.3
%%%%%%%%%%%%%%%%%%%%%%%%%%%%%%%%%%%%%%%%%%%%%%%%%%%%%%%%%%%%%%%%%%%
% Morris GSA
\begin{figure}[h!]
    \centering
    \includegraphics[width=\textwidth]{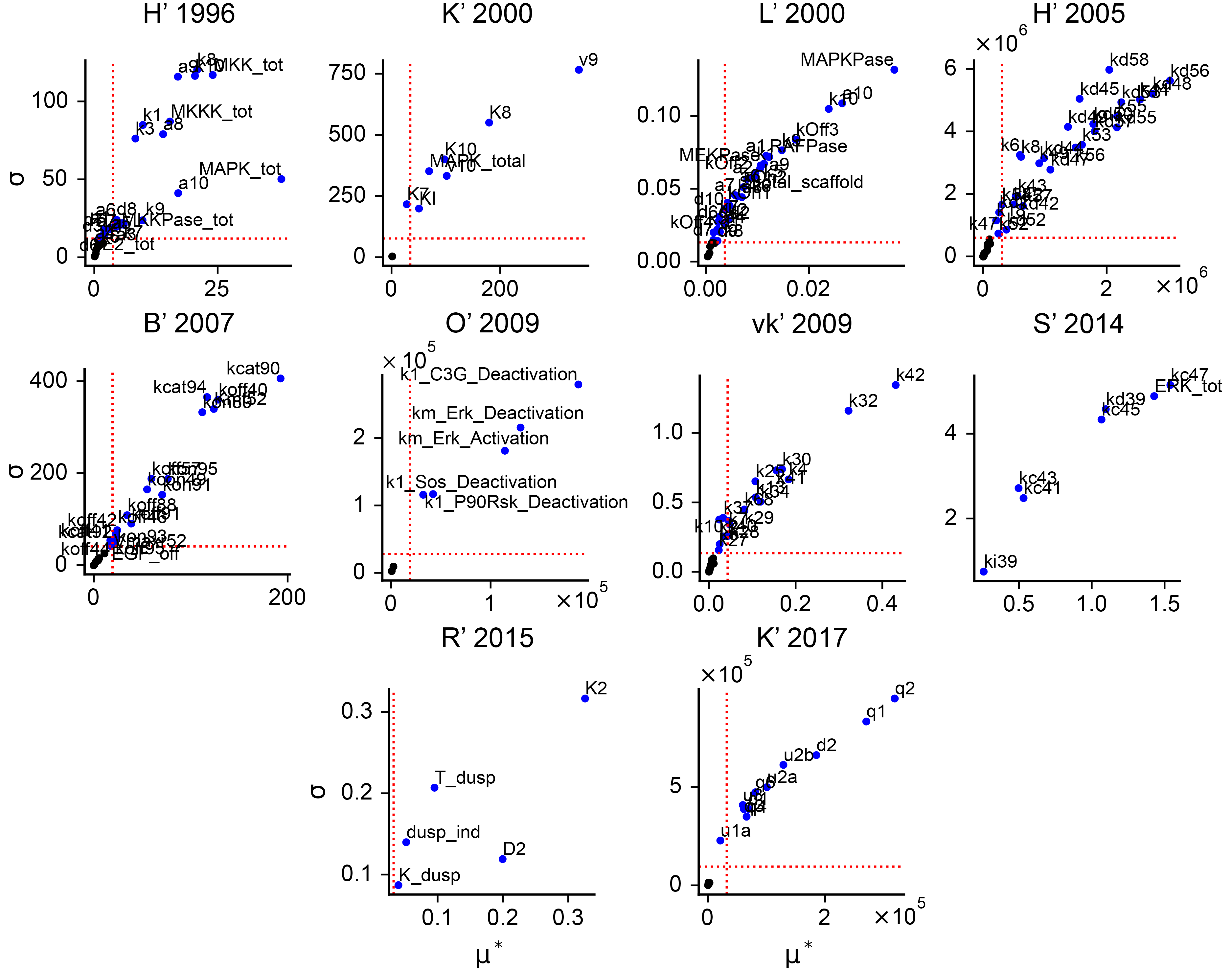}
    \caption{\textbf{Morris sensitivity analysis reveals the parameters of ten ERK signaling models that most strongly influence ERK activity.}
    Scatter plots of Morris effects $\sigma$ and $\mu^*$ normalized to the maximum of each effect independently for all models.
    Blue points with labels show \textit{influential} parameters determined by $\sigma_i \leq 0.1$ or $\mu^*_i \geq 0.1$ for each parameter $\theta_i$.
    Black points are \textit{noninfluential} parameters that do not meet the thresholds.}
    \label{sup-fig:gsa}
\end{figure}

%%%%%%%%%%%%%%%%%%%%%%%%%%%%%%%%%%%%%%%%%%%%%%%%%%%%%%%%%%%%%%%%%%%
% Supp figs -- for Figure 2
%%%%%%%%%%%%%%%%%%%%%%%%%%%%%%%%%%%%%%%%%%%%%%%%%%%%%%%%%%%%%%%%%%%

% Synthetic dose-response post-pred and MMI
\begin{figure}[h!]
    \centering
    \includegraphics[width=\textwidth]{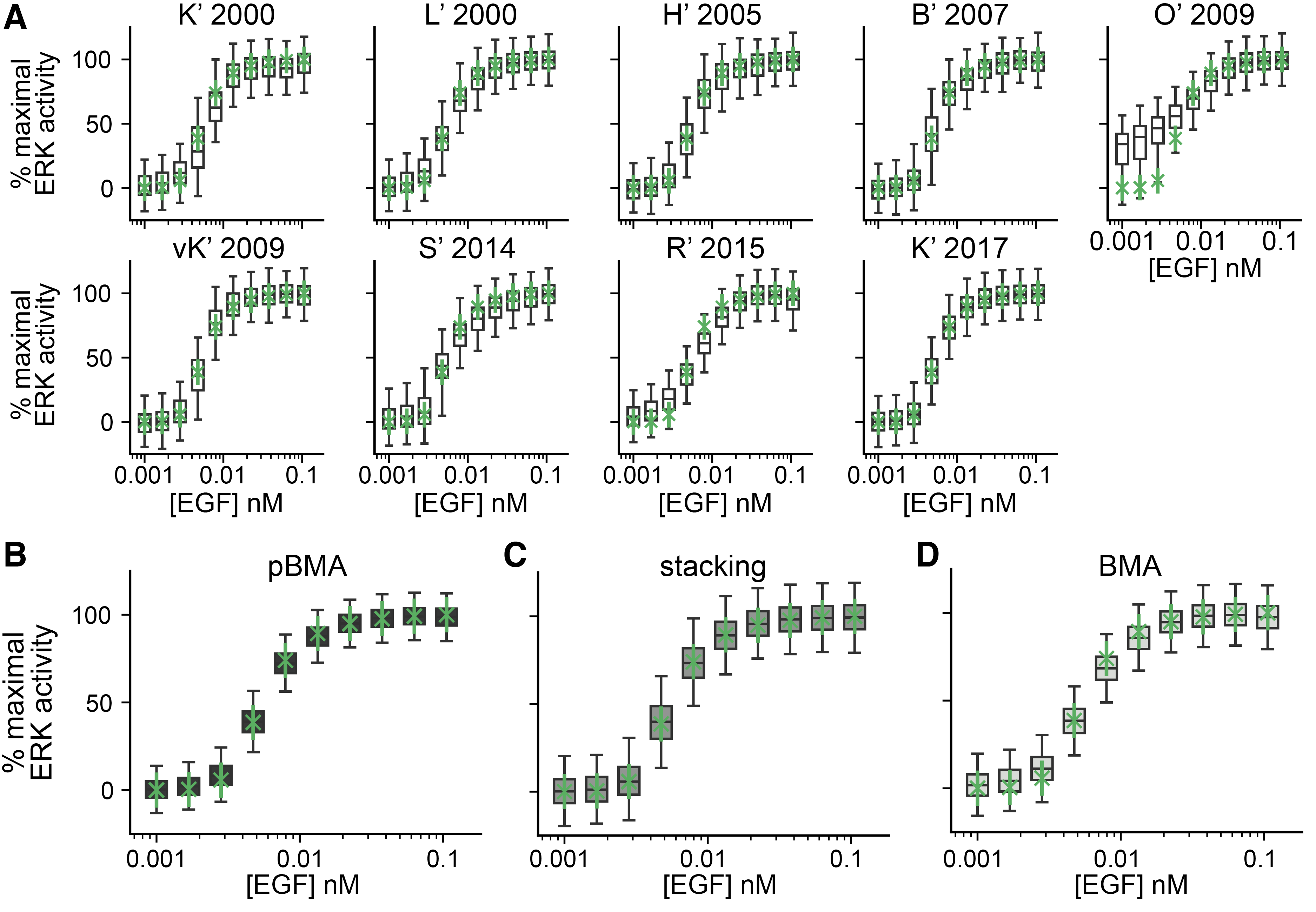}
    \caption{\textbf{Posterior predictive densities of EGF--ERK dose-response curve predictions.}
    (\textbf{A}) Posterior predictive densities of the EGF-ERK dose-response for all models.
    % Box plots constructed with $n=16000$ samples for S'~2014, $n=4000$ samples for K'~2000, R, 2015, and K'~2017, $n=1000$ samples for O'~2009, $n=800$ samples for L'~2000 and $n=400$ for H'~2005, B'~2007, and vK'~2009.
    (\textbf{B})--(\textbf{D}) MMI predictions of the EGF-ERK dose-response curve using pseudo-BMA, stacking, and BMA, respectively.
    }
    \label{sup-fig:synth-DR-post-pred}
\end{figure}

% Synthetic trajectory MMI
\begin{figure}[h!]
    \centering
    \includegraphics[width=\textwidth]{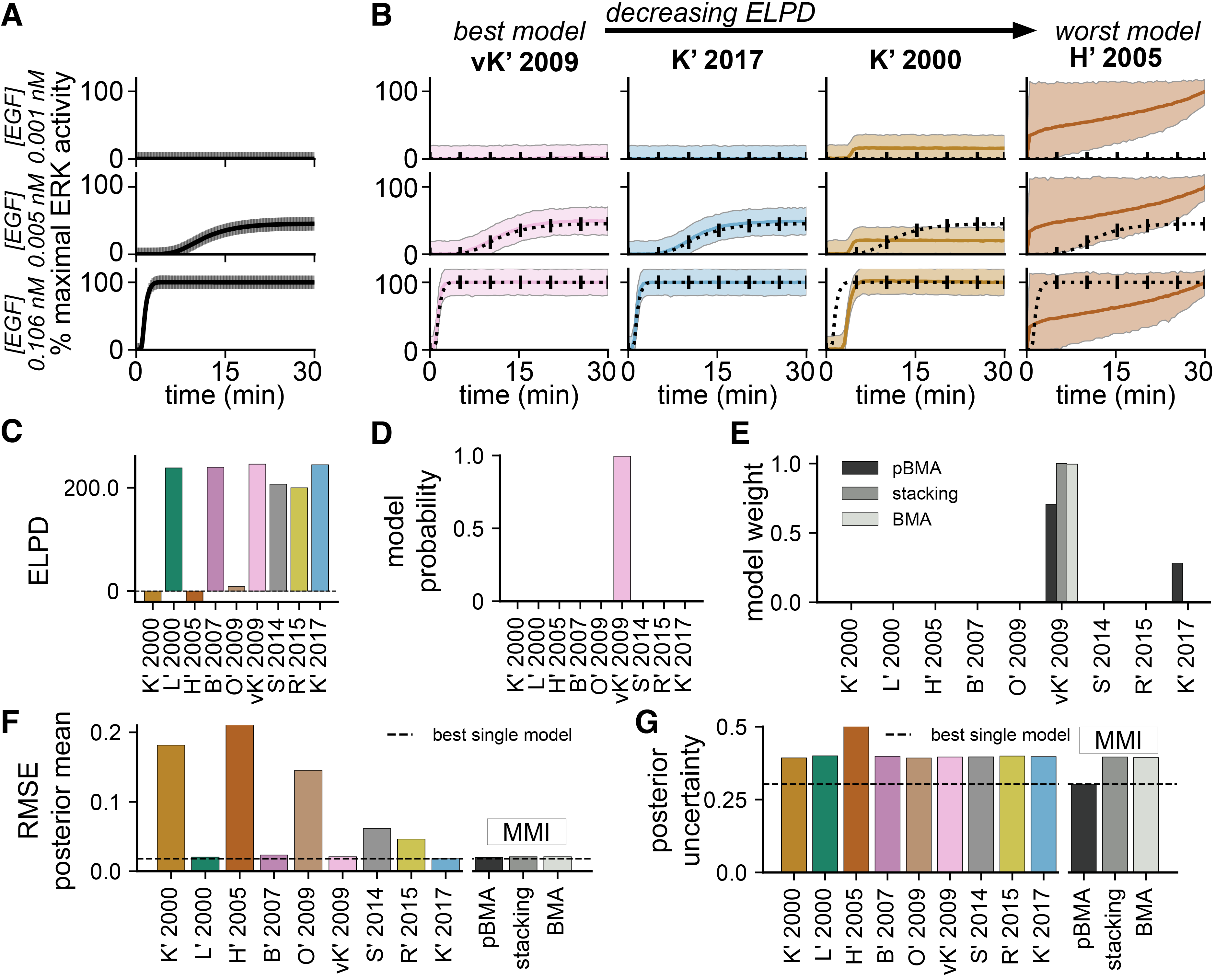}
    \caption{\textbf{Multimodel inference of dynamic EGF--induced ERK activity.}
    (\textbf{A}) Synthetic EGF-induced ERK activity trajectory data generated by simulating the HF' 1996 model~\cite{Huang1996-ki} at three EGF inputs, $0.001$ nM, $0.005$ nM, and $0.106$ nM, for 30 minutes and normalizing the total active ERK concentration to the maximum response across the EGF levels to yield \% maximal ERK activity.
    (\textbf{B}) Posterior predictive densities of ERK activity trajectories for four out of nine models ordered by the expected log pointwise predictive density (ELPD), a quantity used to assess predictive performance.
    (\textbf{C}) ELPD values for all nine models considered for MMI estimated using pareto-smoothed importance sampling leave-one-out cross-validation (PSIS-LOO-CV).
    (\textbf{D}) The log model probability of each model.
    (\textbf{E}) Weights assigned to each model for MMI using pseudo-BMA, BMA, and stacking.
    (\textbf{F}) Root mean square error (RMSE) of the posterior mean ERK activity trajectory prediction for each model and the multimodel predictions.
    (\textbf{G}) Posterior uncertainty in the ERK response measured by the average of the $95\%$ credible interval taken over all EGF levels and all times.
    }
    \label{sup-fig:synth-traj}
\end{figure}

% Synthetic trajectory post-pred and MMI
\begin{figure}[h!]
    \centering
    \includegraphics[width=\textwidth]{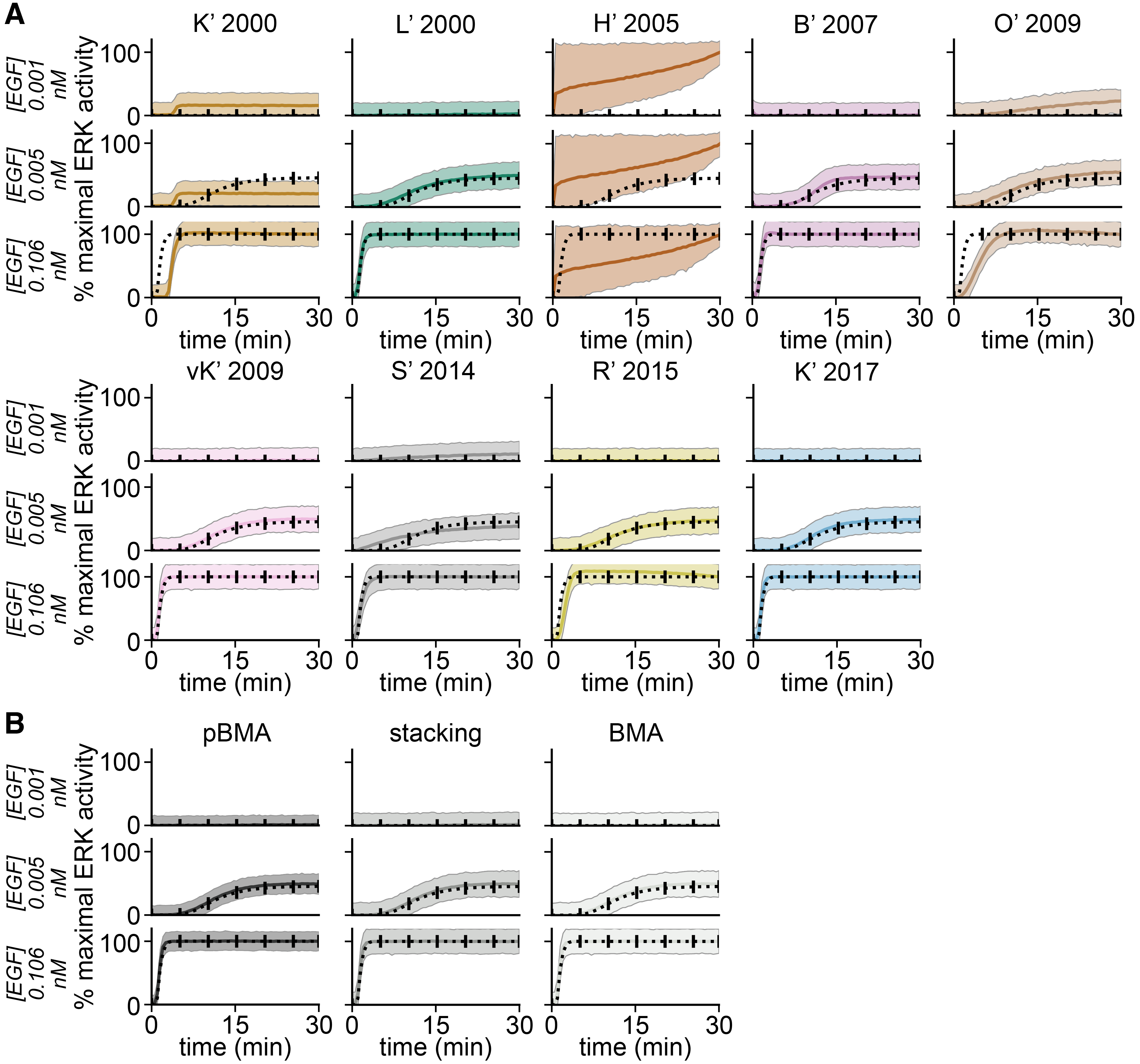}
    \caption{\textbf{Posterior predictive densities of EGF-induced dynamic ERK activity predictions.}
    (\textbf{A}) Posterior predictive densities of trajectories of EGF-induced ERK activity for all models.
    Black dashed line shows the data with error bars indicating the standard deviation.
    Solid colored line shows the posterior predictive mean trajectory.
    Shaded band shows the 95\% credible interval of the posterior predictive density.
    (\textbf{B}) MMI predictions of the EGF-induced ERK activity using pseudo-BMA, stacking, and BMA.
    }
    \label{sup-fig:synth-traj-post-pred}
\end{figure}

%%%%%%%%%%%%%%%%%%%%%%%%%%%%%%%%%%%%%%%%%%%%%%%%%%%%%%%%%%%%%%%%%%%
% Supp figs for Figure 3
%%%%%%%%%%%%%%%%%%%%%%%%%%%%%%%%%%%%%%%%%%%%%%%%%%%%%%%%%%%%%%%%%%%
% fraction models supp
\begin{figure}
    \centering
    \includegraphics[width=\textwidth]{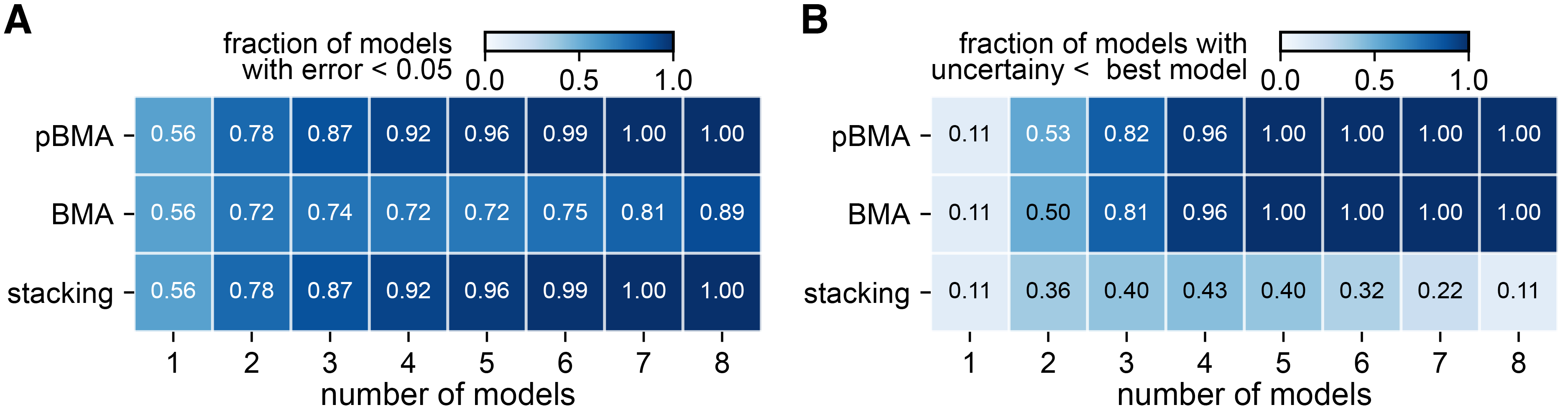}
    \caption{(\textbf{A}) Fraction of MMI estimates with relative error less than 0.05 for MMI estimators with all possible combinations of models at each model set size.
    (\textbf{B}) Fraction of MMI estimates with posterior uncertainty less than that of the best model, R' 2015.}
    \label{sup-fig:synth-mmi-DR-modelCombinotorics_quant}
\end{figure}

% trajectory model perturbation figure
\begin{figure}[h!]
    \centering
    \includegraphics[width=\textwidth]{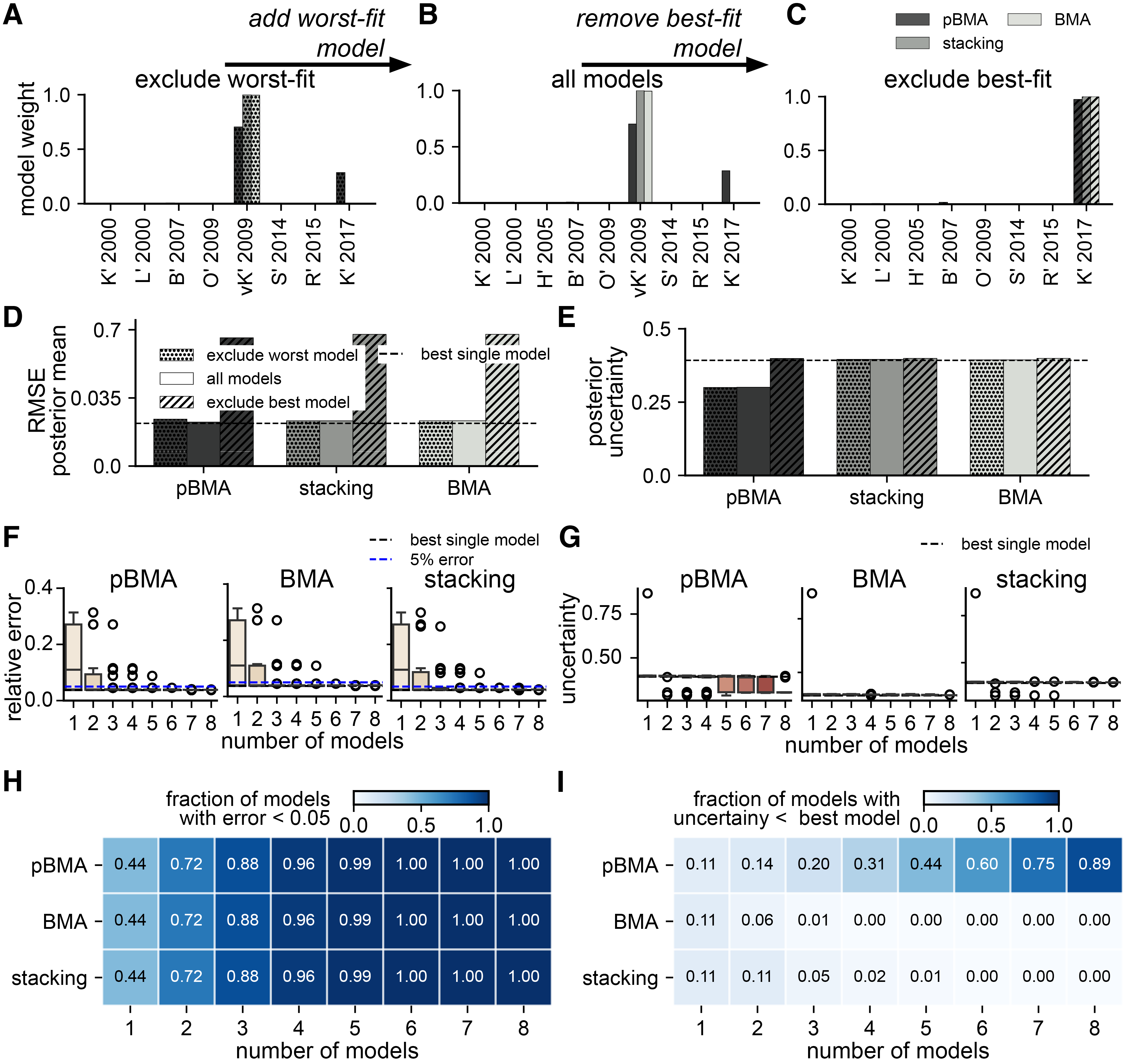}
    \caption{
        \textbf{Bayesian multimodel inference is robust to perturbations in the set of plausible models for dynamic ERK activity predictions.}
        \textit{Effects of adding a bad model or removing a good model.}
        (\textbf{A})--(\textbf{C}) Weights assigned to models in three model sets: (A) excluding the worst-fit model, H'~2005, (B) all models, (C) excluding the best-fit, vK'~2009.
        (\textbf{D}) Root mean square error (RMSE) of the posterior mean dose-response prediction for each MMI method and model set.
        Dotted patterning indicates the set of models without the worst-fit model, no patterning is the complete set, and dashed patterning is the set without the best-fit model.
        The dashed horizontal line is the RMSE of the best-fit model, vK'~2009.
        (\textbf{E}) Average posterior uncertainty in the ERK response measured by the mean of the $95\%$ credible interval taken over all EGF levels.
        \textit{Effects of changing the number of models for MMI.}
        (\textbf{F}) Box plots of relative error of the posterior mean ERK response averaged over all times for MMI predictions with increasing numbers of models.
        Dashed blue line shows 5 \% relative error (0.05), and the dashed black line shows the error of the best model, K'~2017.
        Open circles show outliers.
        (\textbf{G}) Posterior uncertainty of ERK response averaged over time for MMI predictions with increasing numbers of models.
        The dashed black line shows the error of the best model, vK'~2009.
        Open circles show outliers.
        (\textbf{H}) Fraction of MMI estimates with relative error less than 0.05 for MMI estimators with all possible combinations of models at each model set size.
        (\textbf{I}) Fraction of MMI estimates with posterior uncertainty less than that of the best model, O' 2009.
    }
    \label{sup-fig:synth-mmi-traj-perturb}
\end{figure}

%%%%%%%%%%%%%%%%%%%%%%%%%%%%%%%%%%%%%%%%%%%%%%%%%%%%%%%%%%%%%%%%%%%
% Supp figs for Figure 4
%%%%%%%%%%%%%%%%%%%%%%%%%%%%%%%%%%%%%%%%%%%%%%%%%%%%%%%%%%%%%%%%%%%
\begin{figure}[h!p]
    \centering
    \includegraphics[width=\textwidth]{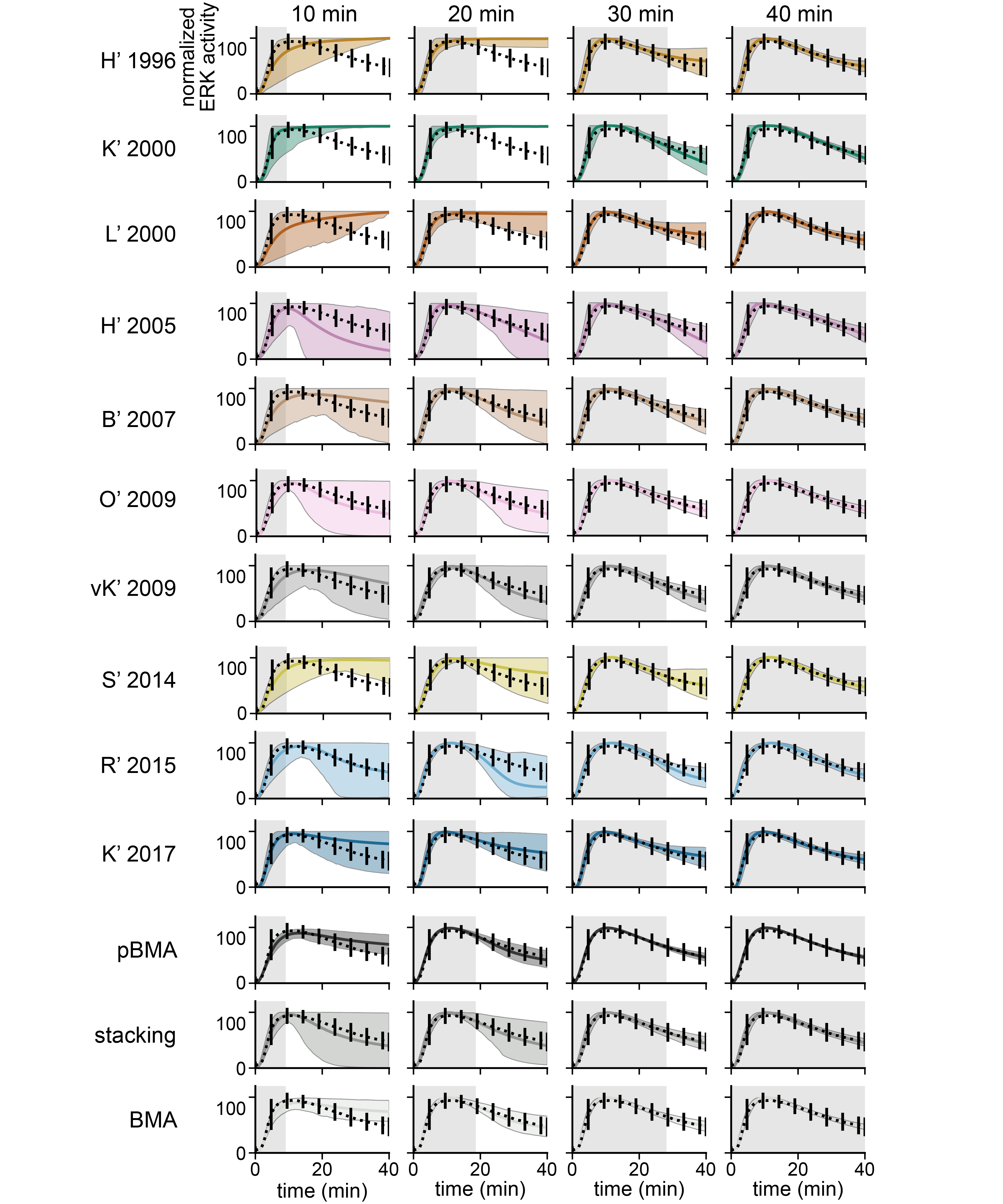}
    \caption{\textbf{Posterior predictive densities of cytoplasmic ERK activity.}
    Black dashed line shows the data with error bars indicating the standard deviation.
    Solid colored line shows the posterior predictive mean trajectory.
    Shaded band shows the 95\% credible interval of the posterior predictive density.
    Corresponds to Figure~\ref{fig:keyes-Rap1-NegFeed}.}
    \label{sup-fig:keyes-data-shorten-SUPP}
\end{figure}

\begin{figure}[h!p]
    \centering
    \includegraphics[width=\textwidth]{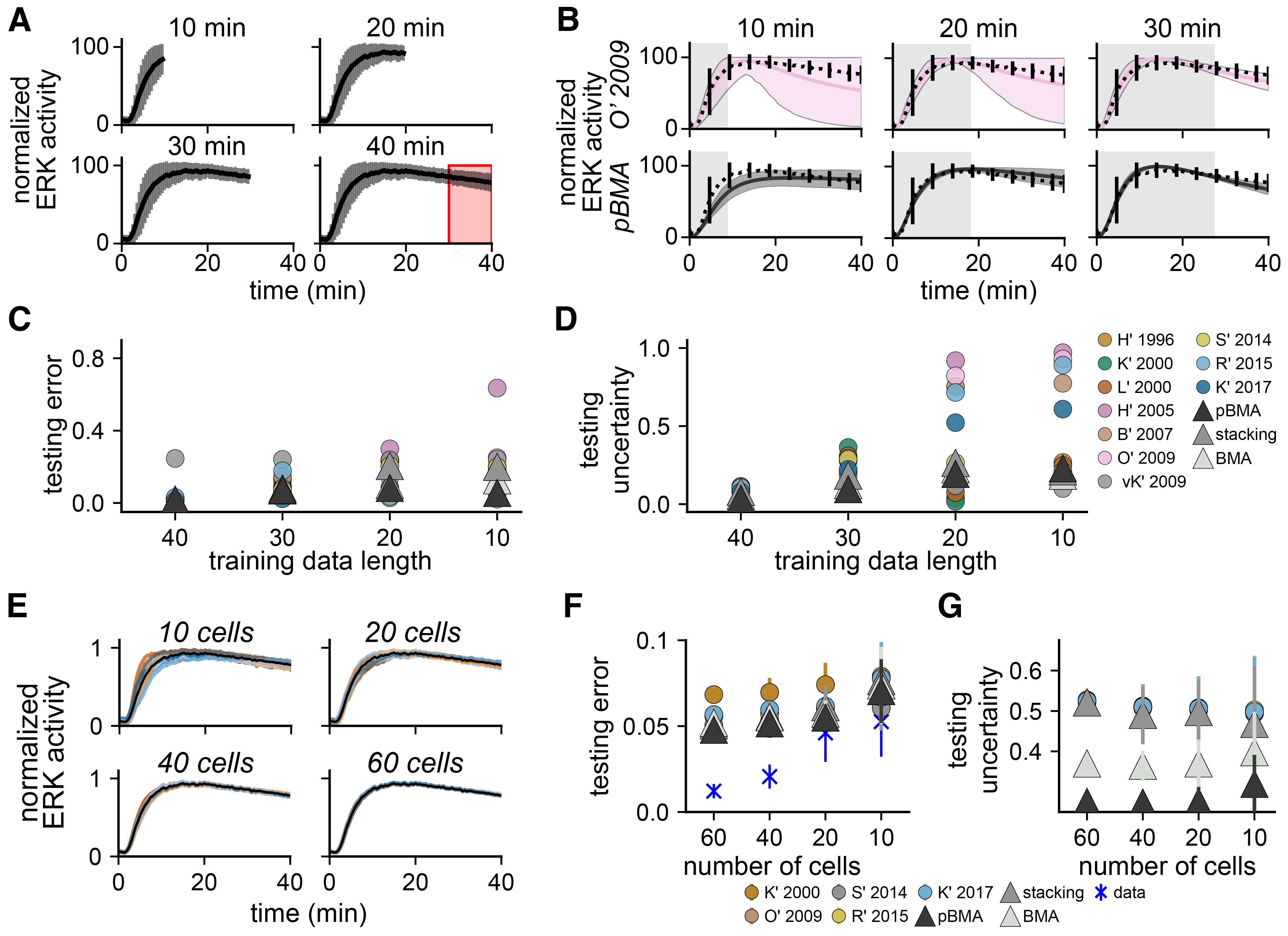}
    \caption{\textbf{Mulitmodel inference predictions of plasma membrane ERK activity are robust to uncertainties due to decreasing data length and quality.}
    \textit{Effects of decreasing data length.}
    (\textbf{A}) Shorter training data was constructed by truncating the original 40-minute plasma membrane ERK trajectories at the 10-, 20-, and 30-minute time points (mean, black trace; and standard deviation, grey bars).
    Predictive performance was assessed by computing errors and uncertainties in the final 10 minutes (red box).
    (\textbf{B}) Posterior predictions from decreased training data for the best model O'~2009 (highest ELPD across all training datasets) and MMI with pseudo-BMA.
    Predictions for additional models and MMI methods are shown in Supplemental Figure~\ref{sup-fig:keyes-data-shorten-SUPP}.
    (\textbf{C}) Predictive error (relative error) for the final 10 minutes ($t=30 \to t=40$ min) of plasma membrane ERK activity.
    (\textbf{D}) Predictive uncertainty (average width of the 95 \% credible interval) for the final 10 minutes of plasma membrane ERK activity.
    (\textbf{E}) Lower-quality training data was generated by averaging over random subsets of 10, 20, 60, and 60 imaged cells.
    The black trace shows an original average of 76 cells.
    Colored traces show averages of 40 replicate random subsets.
    (\textbf{F}) The predictive error (relative error) of plasma membrane ERK activity with lower quality data compared to average activity trajectory using all cells.
    (\textbf{G}) Predictive uncertainty of plasma membrane ERK activity with lower quality data.
    (F),(G) Filled circles indicate average error of 40 replicates for individual models and triangles for MMI predictions.
    Error bars show the standard deviation over replicates.
    Blue markers show the error of the raw training data at each subset size compared to the original full-data mean.}
    \label{sup-fig:keyes-qualityPM}
\end{figure}

%%%%%%%%%%%%%%%%%%%%%%%%%%%%%%%%%%%%%%%%%%%%%%%%%%%%%%%%%%%%%%%%%%%
% Supp figs for Figure 4
%%%%%%%%%%%%%%%%%%%%%%%%%%%%%%%%%%%%%%%%%%%%%%%%%%%%%%%%%%%%%%%%%%%
\begin{figure}[h!]
    \centering
    \includegraphics[width=\textwidth]{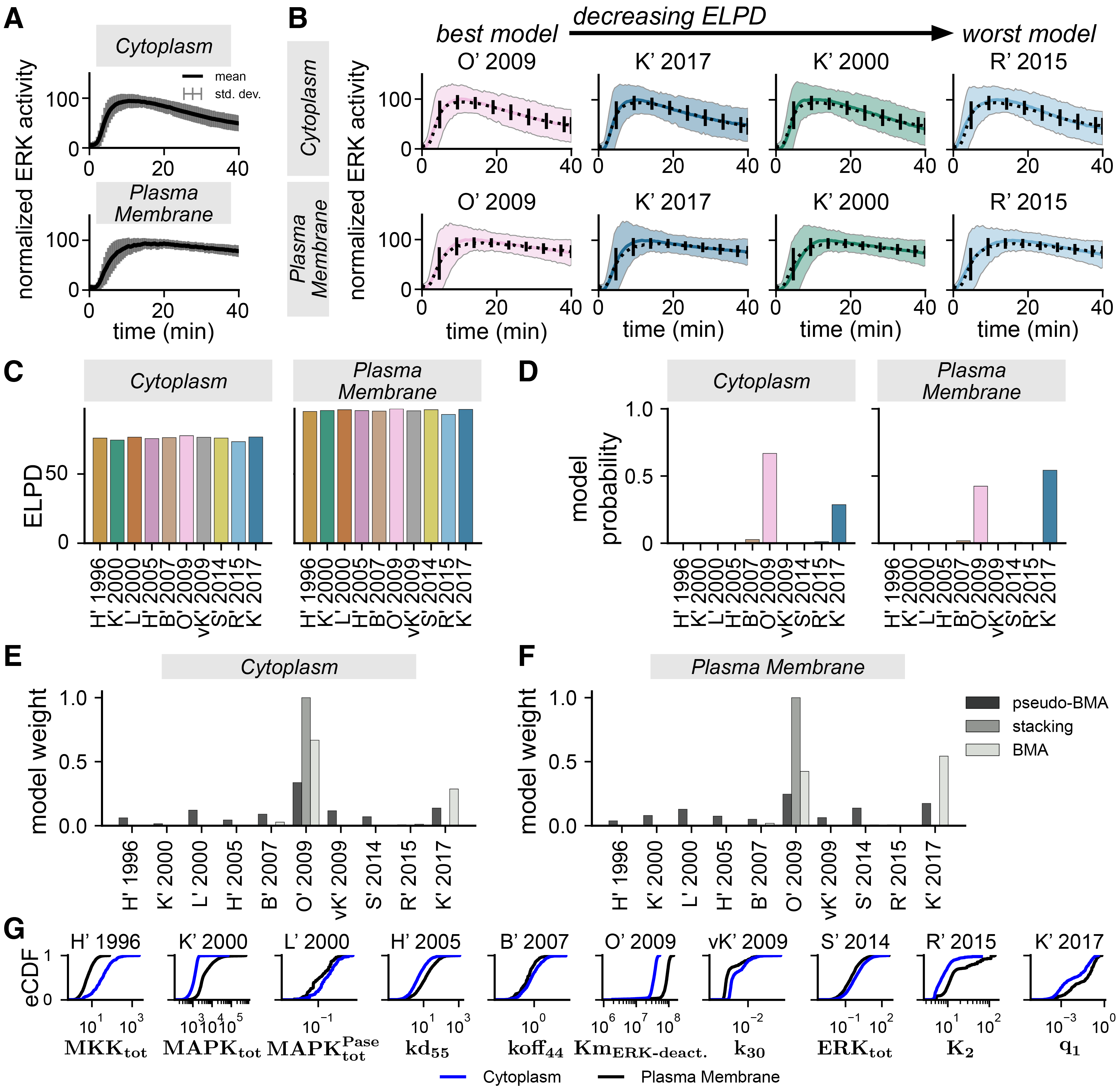}
    \caption{\textbf{Differences in ERK model parameters drive sub-cellular location-specific differences in ERK activity.}
    (\textbf{A}) Normalized sub-cellular location-specific ERK activity measurements.
    Mean (black trace) and standard deviation (grey bars) show statistics computed across all single-cell trajectories. 
    Data originally presented in figure 1 of~\cite{Keyes2020-ub}.
    (\textbf{B}) Posterior predictive trajectories of normalized ERK activity for a subset of the 10 ERK models.
    Models are ordered by decreasing ELPD.
    (\textbf{C}) Estimates of the expected log pointwise predictive density (ELPD) for cytoplasmic and plasma membrane ERK activity.
    (\textbf{D}) Model probabilities in the cytoplasm and plasma membrane.
    (\textbf{E})--(\textbf{F}) Weight assigned to each model by MMI methods in the cytoplasm (E) and plasma membrane (F).
    (\textbf{G}) Estimated empirical cumulative density functions (eCDF) for the model parameters that showed the greatest variation between locations.
    All densities are statistically significantly different between locations ($p< 0.05$ by the Mann-Whitney U-test with a two-sided hypothesis).
    Blue indicates the cytoplasm, and black indicates the plasma membrane.
    }
    \label{sup-fig:keyes-comp1}
\end{figure}

\begin{figure}[h!p]
    \centering
    \includegraphics[width=\textwidth]{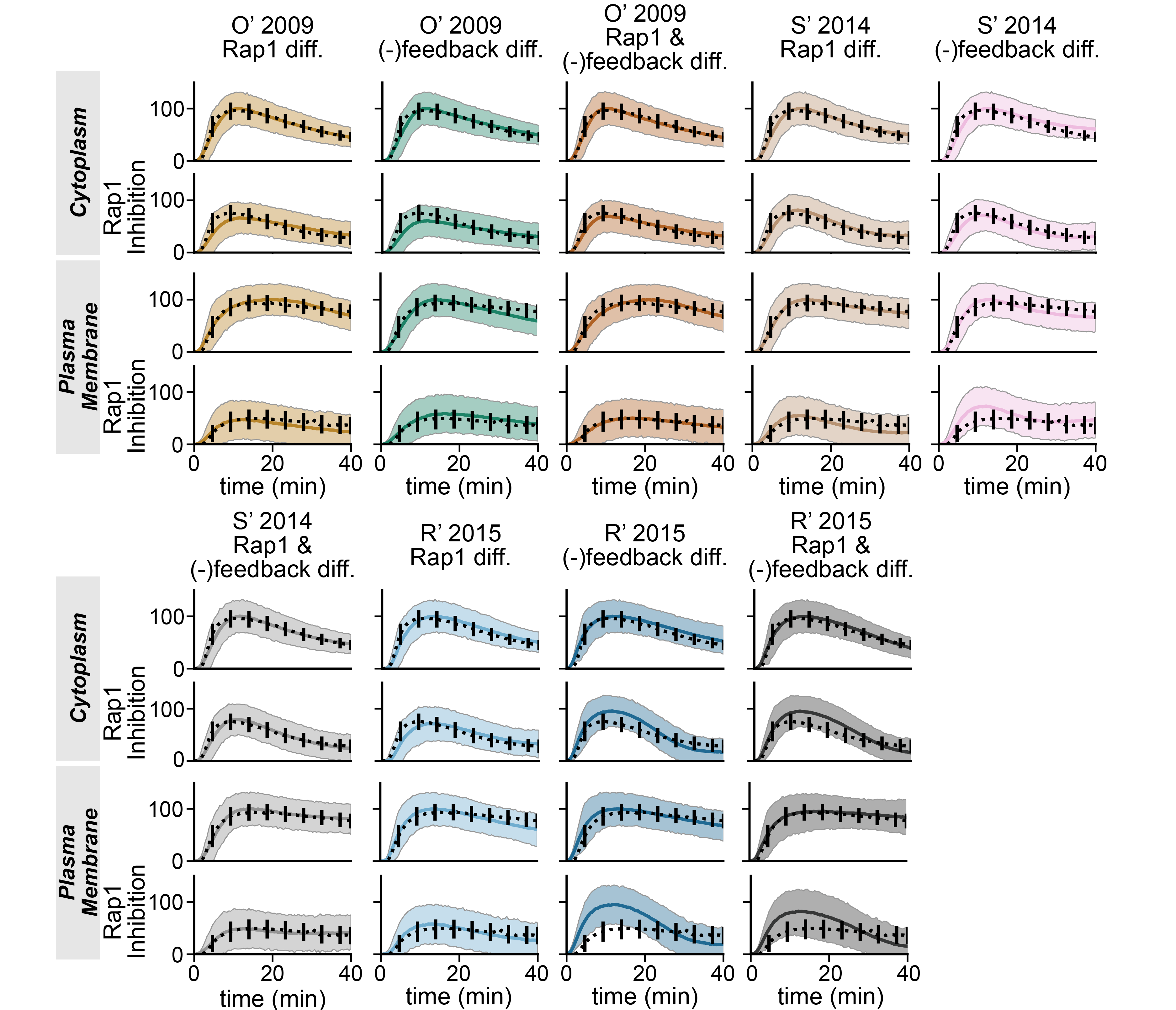}
    \caption{\textbf{Posterior predictive densities of ERK activity under different location-specific hypotheses.}
    Black dashed line shows the data with error bars indicating the standard deviation.
    Solid colored line shows the posterior predictive mean trajectory.
    Shaded band shows the 95\% credible interval of the posterior predictive density.
    Corresponds to Figure~\ref{fig:keyes-Rap1-NegFeed}.}
    \label{sup-fig:keyes-locDiffSUPP}
\end{figure}

%%%%%%%%%%%%%%%%%%%%%%%%%%%%%%%%%%%%%%%%%%%%%%%%%%%%%%%%%%%%%%%%%%%
%%%%%%%%%%%%%%%%%%%%%%%%%%%%%%%%%%%%%%%%%%%%%%%%%%%%%%%%%%%%%%%%%%%
\newpage
 \section{Supplemental Tables} \label{sec:supp-tabs}
 %%%%%%%%%%%%%%%%%%%%%%%%%%%%%%%%%%%%%%%%%%%%%%%%%%%%%%%%%%%%%%%%%%%
%%%%%%%%%%%%%%%%%%%%%%%%%%%%%%%%%%%%%%%%%%%%%%%%%%%%%%%%%%%%%%%%%%%
 % Table of MAPK parameter and state number info
\begin{table}[h!]
    \centering
    \caption{\textbf{Set of extracellular signal-regulated kinase (ERK) signaling models used for Bayesian multimodel inference.}
    The number of state variables (\# states ($n_k$)) represents the number of biochemical species in the model and the number of parameters (\# parameters ($p_k$)) is the total number of free model parameters.
    The number of locally identifiable (\# locally ID) is the number of parameters kept free after local identifiability analysis, and the number of sensitive parameters (\# sensitive) is the number of parameters kept free after global sensitivity analysis.
    }
    \begin{tabular}{ccccc}
        \toprule
        % \hline
        \textbf{Model Name} & \textbf{\# states ($n_k$)} & \textbf{\# parameters ($p_k$)} & \textbf{\# locally ID} & \textbf{\# sensitive} \\
        \midrule
        % \hline
        % \hline
        H'~1996~\cite{Huang1996-ki}          & 16 & 36 & 36 & 26 \\ %\hline
        K'~2000~\cite{Kholodenko2000-ot}      & 6 & 25 & 8 & 6 \\ %\hline
        L'~2000~\cite{Levchenko2000-gg}       & 27 & 40 & 40 & 25 \\ %\hline
        H'~2005~\cite{Hornberg2005-fs}        & 103 & 96 & 94 & 28 \\ %\hline
        B'~2007~\cite{Birtwistle2007-dw}      & 117 & 216 & 90 & 20 \\ %\hline
        O'~2009~\cite{Orton2009-rd}           & 26 & 59 & 7 & 5 \\ %\hline
        vK'~2009~\cite{Von_Kriegsheim2009-nq} & 37 & 77 & 31 & 18 \\ %\hline
        S'~2014~\cite{Shin2014-nd}            & 7 & 20 & 7 & 7 \\ %\hline
        R'~2015~\cite{Ryu2015-ix}             & 17 & 45 & 5 & 5 \\ %\hline
        K'~2017~\cite{Kochanczyk2017-jc}      & 35 & 23 & 14 & 11 \\
        \bottomrule
    \end{tabular}
    \label{tab:ERK-model-info}
\end{table}

% ODE solver hyperparams
\begin{table}[h!]
    \centering
    \caption{\textbf{ODE solver and steady-state methods and corresponding hyperparameters for each ERK model.} 
    We either use an ODE-solver or Newton iteration to find the steady-state solution.
    Steady-state ODE tolerances are used to assess the convergence of ODE-based solutions for either approach.
    We do not compute steady-states for the H' 1996 model, so we omit the corresponding hyperparameters.
    ODE-based solutions are solved until steady-state convergence or $t=t_{\rm max}$.}
    \begin{tabular}{lccccc}
        \toprule
        \textbf{Model} & \begin{tabular}[t]{@{}c@{}}\textbf{ODE tols}\\(\verb!atol!; \verb!rtol!)\end{tabular} & \begin{tabular}[t]{@{}c@{}}\textbf{steady-state}\\\textbf{method}\end{tabular} & \begin{tabular}[t]{@{}c@{}}\textbf{steady-state tols}\\(\verb!atol!; \verb!rtol!)\end{tabular} & \textbf{time units} & $\mathbf{t}_{\textbf{max}}$\\
        \midrule
        H' 1996 & \verb!1e-6!; \verb!1e-6! & -- & -- & sec & -- \\
        K'~2000 & \verb!1e-6!; \verb!1e-6! & Newton & \verb!1e-6!; \verb!1e-6! & sec & $\infty$ \\
        L'~2000 & \verb!1e-6!; \verb!1e-6! & ODE & \verb!1e-10!; \verb!1e-10! & sec & 100{,}000 (s) \\
        H'~2005 & \verb!1e-6!; \verb!1e-6! & ODE & \verb!1e-10!; \verb!1e-10! & sec & 12{,}000 \\
        B'~2007 & \verb!1e-6!; \verb!1e-6! & ODE & \verb!1e-5!; \verb!1e-6! & sec & 10{,}000 \\
        O'~2009 & \verb!1e-6!; \verb!1e-6! & ODE & \verb!1e-5!; \verb!1e-6! & min & 1{,}400 \\
        vK'~2009 & \verb!1e-6!; \verb!1e-6! & ODE & \verb!1e-5!; \verb!1e-6! & sec & 5{,}400  \\
        S'~2014 & \verb!1e-6!; \verb!1e-6! & Newton & \verb!1e-6!; \verb!1e-6! & min & $\infty$ \\
        R'~2015 & \verb!1e-6!; \verb!1e-6! & Newton & \verb!1e-5!; \verb!1e-5! & min & 540\\
        K'~2017 & \verb!1e-6!; \verb!1e-6! & Newton & \verb!1e-6!; \verb!1e-6! & sec & 10{,}800 \\
        \bottomrule
    \end{tabular}
    \label{tab:ODE-hyperparam}
\end{table}

%%%%% SMC stats and info for each set of inferences %%%%%
\begin{table}[h!]
    \centering
    \caption{\textbf{Sampling runtimes and sample counts for all Sequential Monte Carlo sampling.}
    Wall-clock times are shown in hours.
    Parentheses show: (\# of chains and \# number of samples/chain) if those values differ from: \# of chains = 4 and \# number of samples/chain = 500, i.e. (4; 500)}
    \begin{tabular}{lllll}
        \toprule
        \textbf{Model} & Figure~\ref{fig:synth-DR} & Supp. Figure~\ref{sup-fig:synth-traj} & Figure~\ref{fig:keyes-Rap1-NegFeed} & Supp. Fig~\ref{sup-fig:keyes-comp1} \\
        \midrule
        H' 1996 &   -                              & -                              & - & $\mathbf{10.68}$               \\
        K'~2000  & $\mathbf{1.72  }$ ($4$; $1000$) & $\mathbf{7.56 }$               & - & $\mathbf{0.04 }$            \\
        L'~2000  & $\mathbf{8.65  }$ ($4$; $200$)  & $\mathbf{0.66 }$               & - & $\mathbf{0.07 }$            \\
        H'~2005  & $\mathbf{110.66}$ ($4$; $100$)  & $\mathbf{12.13}$ ($4$; $35$)   & - & $\mathbf{1.97 }$ ($8$; $100$) \\
        B'~2007  & $\mathbf{167.87}$ ($4$; $100$)  & $\mathbf{55.21}$               & - & $\mathbf{1.73 }$            \\
        O'~2009  & $\mathbf{2.10  }$ ($4$; $250$)  & $\mathbf{0.12 }$ ($4$; $100$)  & - & $\mathbf{0.11 }$               \\
        vK'~2009 & $\mathbf{0.86  }$ ($4$; $100$)  & $\mathbf{0.48 }$               & - & $\mathbf{0.31 }$            \\
        S'~2014  & $\mathbf{16.90 }$ ($4$; $1000$) & $\mathbf{0.32 }$ ($4$; $1000$) & - & $\mathbf{0.07 }$             \\
        R'~2015  & $\mathbf{18.64 }$ ($4$; $1000$) & $\mathbf{0.05 }$               & - & $\mathbf{0.04 }$            \\
        K'~2017  & $\mathbf{222.76}$ ($4$; $1000$) & $\mathbf{1.83 }$               & - & $\mathbf{0.19 }$            \\
        \begin{tabular}[t]{@{}l@{}}O'~2009\\~~~Rap1 diff.\end{tabular}                 & - & - & $\mathbf{0.24}$ & - \\
        \begin{tabular}[t]{@{}l@{}}O'~2009\\~~~(-)FB diff.\end{tabular}          & - & - & $\mathbf{0.20}$ & - \\
        \begin{tabular}[t]{@{}l@{}}O'~2009\\~~~Rap1 and (-)FB diff.\end{tabular} & - & - & $\mathbf{0.23}$ & - \\
        \begin{tabular}[t]{@{}l@{}}S'~2014\\~~~Rap1 diff.\end{tabular}                 & - & - & $\mathbf{0.06}$ & - \\
        \begin{tabular}[t]{@{}l@{}}S'~2014\\~~~(-)FB diff.\end{tabular}          & - & - & $\mathbf{0.05}$ & - \\
        \begin{tabular}[t]{@{}l@{}}S'~2014\\~~~Rap1 and (-)FB diff.\end{tabular} & - & - & $\mathbf{0.06}$ & - \\
        \begin{tabular}[t]{@{}l@{}}R'~2015\\~~~Rap1 diff.\end{tabular}                 & - & - & $\mathbf{0.15}$ & - \\
        \begin{tabular}[t]{@{}l@{}}R'~2015\\~~~(-)FB diff.\end{tabular}          & - & - & $\mathbf{0.13}$ & - \\
        \begin{tabular}[t]{@{}l@{}}R'~2015\\~~~Rap1 and (-)FB diff.\end{tabular} & - & - & $\mathbf{0.11}$ & - \\
        \bottomrule
    \end{tabular}
    \label{tab:SMC-runtime}
\end{table}

\section{Supplemental Text} \label{sec:supp-text}

\subsection*{Bayesian parameter estimation for sub-cellular location-specific ERK activity} \label{sup-sec:loc-ERK-keyes}

Initially, we hypothesized that location-specific probability densities for model parameters can explain location-specific ERK activity observed by Keyes et al.~\cite{Keyes2020-ub}. 
To test this, we estimated ERK signaling model parameters from observations of cytoplasmic and plasma membrane ERK activity independently (Figure~\subpanelref{sup-fig:keyes-comp1}{A}).
To enable comparison with model predictions, we normalized single-cell EKAR4 YFP/CFP emission ratios and model predictions to be between zero and one.
The resulting posterior predictions of ERK activity show that all models appeared to capture both cytoplasmic and plasma membrane ERK activity equally well, possibly because the estimated probability densities for the parameters varied between intracellular locations (Figure~\subpanelref{sup-fig:keyes-comp1}{B}).
Quantitatively, the predicted ELPDs and model probability are similar across the models within each location (Supplemental Figure~\subpanelref{sup-fig:keyes-comp1}{C}--~\subpanelref{sup-fig:keyes-comp1}{D}).
These findings suggest that the model structure is mostly unimportant for predicting location-specific differences in ERK activation dynamics when all model parameters vary independently between the cytoplasm and the plasma membrane.

Interestingly, the estimated posterior densities confirm that each model predicts different sets of location-specific parameters that drive sub-cellular variations in ERK activity.
Figure~\subpanelref{sup-fig:keyes-comp1}{G} shows the estimated empirical cumulative density functions for the parameters that vary most significantly between locations for each model.
However, a closer examination of the reactions corresponding to parameters with significant differences between locations reveals no clear trends in the parameter estimates (data not shown).
For example, the equilibrium coefficient for the ERK dephosphorylation reaction varied significantly in five of the models; however, two of them predicted stronger ERK dephosphorylation in the cytoplasm than in the plasma membrane, while the other three predicted weaker cytoplasmic ERK dephosphorylation.
Additionally, many of the predicted differences are not physiologically meaningful, e.g., kinase deactivation rates that differ between sub-cellular locations.
Only one model in particular, O'~2009, showed differences in parameters that control Rap1 activation, which was consistent with experimental observations~\cite{Keyes2020-ub}.
Therefore, we concluded that allowing every estimated model parameter to vary independently between sub-cellar locations yields accurately predicts sub-cellular variability in ERK activity without adhering to physiological reality.

\section{Supplemental Materials} \label{sec:supp-mat}

We provide an Excel file that includes details about the nominal parameters and initial conditions for all 10 ERK signaling models used in this work.
The file is available for download from our GitHub repository at this \href{https://github.com/RangamaniLabUCSD/multimodel-inference/blob/main/results/model_info_supplemental_material.xlsx}{link}.
\end{appendix}

\end{document}